\documentclass[reprint,
 superscriptaddress,
nofootinbib,
 amsmath,amssymb,
 aps,
pra,
]{revtex4-2}

\usepackage[utf8]{inputenc} 
\usepackage[T1]{fontenc}    
\usepackage{hyperref}       
\usepackage{url}            
\usepackage{booktabs}       
\usepackage{amsfonts}       
\usepackage{nicefrac}       
\usepackage{microtype}      
\usepackage{graphicx}
\usepackage{xcolor, etoolbox}
\usepackage{amsmath,amsfonts,amsthm} 
\usepackage{mathtools}
\usepackage{ragged2e}
\usepackage{braket}      
\usepackage[english]{babel} 
\graphicspath{{./figures/}} 
\usepackage{enumitem}
\usepackage{xspace}
\usepackage{dsfont}
\usepackage{bm}
\usepackage{bbm}
\usepackage{dcolumn}
\usepackage{mathrsfs}
\usepackage{appendix} 

\newcommand{\Tr}{\text{Tr}}

\hypersetup{
    colorlinks,
    linkcolor=blue,
    citecolor=blue,
    urlcolor=blue,
    breaklinks=true 
}
\DeclarePairedDelimiter\abs{\lvert}{\rvert}

\begin{document}

\title{Accessing continuous-variable entanglement witnesses with multimode spin observables}

\author{Célia Griffet}
    \email{celia.griffet@ulb.be}
    \affiliation{Centre for Quantum Information and Communication, \'Ecole polytechnique de Bruxelles, CP 165, Universit\'e libre de Bruxelles, 1050 Brussels, Belgium}
\author{Tobias Haas}
    \email{tobias.haas@ulb.be}
    \affiliation{Centre for Quantum Information and Communication, \'Ecole polytechnique de Bruxelles, CP 165, Universit\'e libre de Bruxelles, 1050 Brussels, Belgium}
\author{Nicolas J. Cerf}
    \email{nicolas.cerf@ulb.be}
    \affiliation{Centre for Quantum Information and Communication, \'Ecole polytechnique de Bruxelles, CP 165, Universit\'e libre de Bruxelles, 1050 Brussels, Belgium}

\begin{abstract}
We present several measurement schemes for accessing separability criteria for continuous-variable bipartite quantum systems. Starting from moments of the bosonic mode operators, criteria suitable to witness entanglement are expressed in terms of multimode spin observables via the Jordan-Schwinger map. These observables are typically defined over a few replicas of the state of interest and can be transformed into simple photon-number measurements by passive optical circuits. Our measurement schemes require only a handful of measurements, thereby allowing one to efficiently detect entanglement without the need for costly state tomography as illustrated for a variety of physically relevant states (Gaussian, mixed Schrödinger cat, and NOON states). The influence of typical experimental imperfections is shown to be moderate.
\end{abstract}

\maketitle

\section{Introduction}
\label{sec:Introduction}
When analyzing the correlations between two systems, the dividing line between the classical and the quantum is marked by the phenomenon of entanglement. Over the last two decades, plenty of methods for characterizing entanglement theoretically as well as experimentally have been put forward \cite{Plenio2007,Horodecki2009,Guehne2009}. A common strategy relies on demonstrating entanglement by violating a set of experimentally accessible conditions fulfilled by all separable states and violated by a few entangled ones. This includes the prominent Peres-Horodecki (PPT) criterion \cite{Peres1996,Horodecki1996}, which states that bipartite entanglement can be certified when the partially transposed density operator exhibits a negative eigenvalue.

Conditions implied by this PPT criterion have been studied extensively in the framework of continuous variable quantum systems \cite{Braunstein2005b,Weedbrook2012,Serafini2017}, where entanglement detection is further complicated by the infinite dimensional Hilbert space. This encompasses formulations based on uncertainty relations for second moments \cite{Duan2000,Simon2000,Mancini2002,Giovannetti2003,Hertz2016}, fourth-order moments \cite{Agarwal2005}, entropies over canonical variables \cite{Walborn2009,Walborn2011,Saboia2011,Tasca2013,Schneeloch2018}, as well as entropic quantities based on the Husimi $Q$-distribution \cite{Haas2021,Haas2022a,Haas2022b,Haas2022c,Haas2022d}. With these approaches, entanglement could be certified experimentally in the context of quantum optics \cite{Dong2008,Schneeloch2019,Asavanant2019,Qin2019,Moody2022} and with cold atoms \cite{Gross2011,Strobel2014,Peise2015,Fadel2018,Kunkel2018,Lange2018,Kunkel2021}.

Although all of the aforementioned criteria are implied by the PPT criterion, they are generally weaker in the sense that they can not detect entanglement for a few entangled states that have a negative partial transpose. A complete hierarchy of conditions in terms of moments of the bosonic mode operators, being sufficient \textit{and} necessary for the negativity of the partial transpose, has been put forward by Shchukin and Vogel in \cite{Shchukin2005}, and was further developed in Refs. \cite{Miranowicz2006,Miranowicz2009}. While this approach settled the quest for faithfully evaluating the negativity of the partially transposed state, efficient methods for accessing the most important low-order conditions have until now remained elusive. 

In this paper, we put forward simple measurement schemes of these low-order conditions by introducing multimode spin observables which act on a few replicas (\textit{i. e.}, independent and identical copies) of the bipartite state of interest. Contrary to local canonical operators, whose low-order correlation functions have to be measured through costly tomographic routines involving homodyne measurements \cite{Welsch1997,Mancini1997,Cramer2010}, such multimode spin observables can be transformed into a bunch of photon number measurements by using passive optical elements \cite{Hertz2019b,Griffet2022a,Griffet2022b}. Following this multicopy technique, we devise measurement protocols for three of the most interesting separability criteria obtained in Ref. \cite{Shchukin2005} and illustrate how they efficiently witness entanglement for the classes of Gaussian, mixed Schrödinger cat, and NOON states, respectively.
In all cases, we discuss how experimental imperfections may affect the detection capability.

\textit{The remainder of this paper is organized as follows}.
We begin \autoref{sec:Preliminaries} with a brief recapitulation of the Shchukin-Vogel hierarchy for entanglement witnesses (\autoref{subsec:ShchukinVogelHierarchy}), followed by an overview of the multicopy method (\autoref{subsec:MultimodeMethod}) and, specifically, of the Jordan-Schwinger map used to build multimode spin observables  (\autoref{subsec:SpinOperators}). Thereupon, we derive and evaluate multimode expressions for three important classes of entanglement criteria in \autoref{sec:EntanglementCriteria}, that is, criteria that are best suited for Gaussian states (\autoref{subsec:GaussianStates}), mixed Schrödinger cat states (\autoref{subsec:MixedSchrödingerCatStates}), and NOON states (\autoref{subsec:NOONStates}). We also discuss the influence of imperfect preparation and losses for each criterion. Finally, we summarize our findings and provide an outlook in \autoref{sec:Conclusion}.

\textit{Notation}. We employ natural units $\hbar = 1$ and use bold (normal) letters for quantum operators $\boldsymbol{O}$ (classical variables $O$). We write $\braket{\boldsymbol{O}} = \Tr \{ \boldsymbol{\rho} \, \boldsymbol{O} \}$ for single-copy expectation values and $\braket{\dots \braket{\boldsymbol{O}} \dots} = \Tr \{ (\boldsymbol{\rho} \otimes \dots \otimes \boldsymbol{\rho}) \, \boldsymbol{O} \}$ for multicopy expectation values. The modes $\boldsymbol{a}$ and $\boldsymbol{b}$ are associated with Alice's and Bob's subsystems $A$ and $B$, respectively, and copies are labeled by greek indices $\mu, \nu$.

\section{Preliminaries}
\label{sec:Preliminaries}

\subsection{Shchukin-Vogel hierarchy}
\label{subsec:ShchukinVogelHierarchy}
We consider a bipartite continuous-variable quantum system $AB$ with local bosonic mode operators $\boldsymbol{a}$ and $\boldsymbol{b}$ satisfying $[\boldsymbol{a},\boldsymbol{a}^{\dagger}] = [\boldsymbol{b},\boldsymbol{b}^{\dagger}] = 1$ (we restrict to Alice's and Bob's subsystems consisting of one single mode). By the Peres-Horodecki criterion, all separable states have a non-negative partial transpose $\boldsymbol{\rho}^{T_2} \ge 0$ \cite{Peres1996,Horodecki1996}. Following \cite{Shchukin2005,Miranowicz2006,Miranowicz2009}, the non-negativity of the partial transpose $\boldsymbol{\rho}^{T_2}$ can be assessed in full generality by demanding that for all normally-ordered operators $\boldsymbol{f} = \sum_{n,m,k,l} c_{n m k l} \, \boldsymbol{a}^{\dagger n} \boldsymbol{a}^m \boldsymbol{b}^{\dagger k} \boldsymbol{b}^l$ with complex-valued $c$, the inequality $\Tr \left\{ \boldsymbol{\rho}^{T_2} \boldsymbol{f}^{\dagger} \boldsymbol{f} \right\} \ge 0$ is fulfilled, which can be expressed as a bilinear form in $c$ with the matrix of moments $D_{p q r s, n m k l} = \braket{\boldsymbol{a}^{\dagger q} \boldsymbol{a}^p \boldsymbol{a}^{\dagger n} \boldsymbol{a}^m \boldsymbol{b}^{\dagger s} \boldsymbol{b}^r \boldsymbol{b}^{\dagger k} \boldsymbol{b}^l}$ as 
\begin{equation}
     \sum_{\substack{n,m,k,l \\ p,q,r,s}} c^{*}_{p q r s} \, c_{n m k l} \, D_{p q r s, n m k l}^{T_2} \ge 0
     \label{eq:ShchukinVogelCriteria}
\end{equation}
for all $c$'s. By Silvester's criterion, the latter inequality holds true for all $c$'s if and only if all principal minors of the matrix $D^{T_2}$ are non-negative. Using $D_{p q r s, n m k l}^{T_2} = D_{p q k l, n m r s}$, we obtain that $\boldsymbol{\rho}^{T_2}$ is non-negative if and only if all determinants
\begin{equation}
    \hspace{-0.1cm}d = \abs{D^{T_2}} = \begin{vmatrix}
    1 & \braket{\boldsymbol{a}} & \braket{\boldsymbol{a}^{\dagger}} & \braket{\boldsymbol{b}^{\dagger}} & \braket{\boldsymbol{b}} & \dots \\
    \braket{\boldsymbol{a}^{\dagger}} & \braket{\boldsymbol{a}^{\dagger} \boldsymbol{a}} & \braket{\boldsymbol{a}^{\dagger 2}} & \braket{\boldsymbol{a}^{\dagger} \boldsymbol{b}^{\dagger}} & \braket{\boldsymbol{a}^{\dagger} \boldsymbol{b}} & \dots \\
    \braket{\boldsymbol{a}} & \braket{\boldsymbol{a}^2} & \braket{\boldsymbol{a} \boldsymbol{a}^{\dagger}} & \braket{\boldsymbol{a} \boldsymbol{b}^{\dagger}} & \braket{\boldsymbol{a} \boldsymbol{b}} & \dots \\
    \braket{\boldsymbol{b}} & \braket{\boldsymbol{a} \boldsymbol{b}} & \braket{\boldsymbol{a}^{\dagger} \boldsymbol{b}} & \braket{\boldsymbol{b}^{\dagger} \boldsymbol{b}} & \braket{\boldsymbol{b}^2} & \dots \\
    \braket{\boldsymbol{b}^{\dagger}} & \braket{\boldsymbol{a} \boldsymbol{b}^{\dagger}} & \braket{\boldsymbol{a}^{\dagger} \boldsymbol{b}^{\dagger}} & \braket{\boldsymbol{b}^{\dagger 2}} & \braket{\boldsymbol{b} \boldsymbol{b}^{\dagger}} & \dots \\
    \vdots & \vdots & \vdots & \vdots & \vdots & \ddots
    \end{vmatrix}
    \label{eq:ShchukinVogelCriteriaDeterminants}
\end{equation}
are non-negative.

By the latter arguments, the negativity of any principal minor of the matrix $D^{T_2}$ provides a sufficient condition for entanglement, \textit{i. e.}, an entanglement witness. This corresponds to specific choices of $\boldsymbol{f}$ by setting the $c$ coefficients in such a way that one only keeps a subset of rows and corresponding columns in the matrix $D^{T_2}$. For example, we will focus in what follows on the determinant of the matrix consisting of the first, second, and fourth rows and columns of $D^{T_2}$, denoted as $d_{1,2,4}$ in Eq. \eqref{eq:d124}, and will use $d_{1,2,4}<0$ as an entanglement witness. Similarly, we will consider $d_{1,4,9}$ [Eq. \eqref{eq:d149}] and $d_{1,9,13}$ [Eq. \eqref{eq:d1913}].

It is interesting to note that all principal minors of \eqref{eq:ShchukinVogelCriteriaDeterminants} are invariant under arbitrary local rotations 
$\boldsymbol{a}\to e^{-i \theta_a}\boldsymbol{a}$ and
$\boldsymbol{b}\to e^{-i \theta_b}\boldsymbol{b}$ with $\theta_a,\theta_b\in [0,2\pi)$,
since, in every term of these determinants, local annihilation and creation operators appear in equal number, so that all phases cancel termwise. This property will in turn carry over to the derived multicopy observables, implying that all entanglement witnesses will be invariant when Alice or Bob apply a local phase shift. The same is \textit{not} true in general for arbitrary local displacements $\boldsymbol{a}\to \boldsymbol{a} + \alpha$ and
$\boldsymbol{b}\to \boldsymbol{b}+ \beta $ with $\alpha,\beta\in\mathbb{C}$, which can be seen by considering for example the subdeterminant $d_{2,3}$. For certain subdeterminants, however, this invariance is restored, which is for example the case for $d_{1,2,4}$ [Eq. \eqref{eq:d124}]. In that case, the effect of local displacements can be discarded with simple optical circuits involving only beam splitters and we may only consider centered states [in this example, it is  then enough to use $d_{2,4}$ as an entanglement witness, see Eq. \eqref{eq:d24}].

\subsection{Multicopy method}
\label{subsec:MultimodeMethod}
All criteria obtainable by Shchukin-Vogel's approach can be expressed in terms of the non-negativity of a determinant $d$ containing moments, which offers the possibility to write them in terms of expectation values of multimode observables. It is indeed known that 
any $n$-th degree polynomial of matrix elements of a state $\boldsymbol{\rho}$ can be accessed by defining some observable acting on a $n$-copy version of the state, namely $\boldsymbol{\rho}^{\otimes n}$ \cite{Brun2004}. Inspired by this multicopy method, tight uncertainty relations \cite{Hertz2019b} as well as nonclassicality witnesses \cite{Griffet2022a} (see also \cite{Griffet2022b}) have been formulated by devising multicopy observables from determinants similar to \eqref{eq:ShchukinVogelCriteriaDeterminants}. The general scheme is as follows. Given the determinant $d$ of a matrix containing expectation values of mode operators, the corresponding multicopy observable $\boldsymbol{D}$ is obtained by dropping all expectation values, assigning one copy to each row and averaging over all permutations of the copies. By construction, the multicopy expectation value $\braket{ \dots \braket{\boldsymbol{D}} \dots }$ coincides with the determinant $d$.

For measuring these observables, the remaining task is to find suitable optical circuits. We start from the $n$-dimensional extension of mode operators describing subsystem $A$, which reads $[\boldsymbol{a}_{\mu}, \boldsymbol{a}^{\dagger}_{\nu}] = \delta_{\mu \nu}$ with $\mu, \nu \in \{1, \dots, n \}$ and $n$ denoting the number of copies (we will of course use a similar notation for copies of subsystem $B$). In order to transform the measurement of some $n$-mode observable $\boldsymbol{D}$ into simple photon-number measurements, we employ passive linear interferometers, which amounts to applying a unitary  transformation (\textit{i. e.}, a passive Bogoliubov transformation) to the mode operators, namely,
\begin{equation}
    (\boldsymbol{a}_1, ..., \boldsymbol{a}_n)^T \to (\boldsymbol{a}_{1'}, ..., \boldsymbol{a}_{n'})^T = M \, (\boldsymbol{a}_1, ..., \boldsymbol{a}_n)^T.
\end{equation}
The unitary matrix $M$ can be decomposed in terms of two building blocks, the beam splitter
\begin{equation}
    \textrm{BS}_{\mu \nu}(\tau) = \begin{pmatrix}
    \sqrt{\tau} & \sqrt{1-\tau} \\
    \sqrt{1-\tau} & -\sqrt{\tau}
    \end{pmatrix},
    \label{eq:BeamSplitter}
\end{equation}
with transmittivity $\tau \in [0,1]$, and the phase shifter
\begin{equation}
    \textrm{PS}_{\mu}(\theta) = e^{-i \theta},
    \label{eq:PhaseShifter}
\end{equation}
with phase $\theta \in [0, 2\pi)$, where $\mu$ and $\nu$ designate the mode indices on which the corresponding transformations are applied.

\subsection{Multimode spin operators}
\label{subsec:SpinOperators}
When restricting to two modes ($\mu,\nu=1,2$), a particularly useful set of multimode operators can be constructed from algebraic considerations. Considering again subsystem $A$, the fundamental representation of the Lie algebra $su(2)$, \textit{i. e.}, the Pauli matrices $G^j = \sigma^j/2$ fulfilling $[G^j, G^k] = i \, \epsilon_{j k l} \, G^l$ with $j,k,l=1,2,3$, is realized on its two-mode extension by the quantum operators
\begin{equation}
    \boldsymbol{L}^j = \sum_{\mu, \nu} \boldsymbol{a}^{\dagger}_{\mu} \, (G^j)_{\mu \nu} \, \boldsymbol{a}_{\nu} ,
\end{equation}
where $(G^j)_{\mu \nu}$ denotes the $(\mu, \nu)$th entry of the Pauli matrix $G^j$. This is known as the Jordan-Schwinger map. More generally, in the $n$-mode case, this leads to defining the three two-mode spin operators
\begin{equation}
    \begin{split}
        \boldsymbol{L}_{a_{\mu \nu}}^x &= \frac{1}{2} \left( \boldsymbol{a}_{\nu}^{\dagger} \boldsymbol{a}_{\mu} + \boldsymbol{a}_{\mu}^{\dagger} \boldsymbol{a}_{\nu}, \right), \\
        \boldsymbol{L}_{a_{\mu \nu}}^y &= \frac{i}{2} \left( \boldsymbol{a}_{\nu}^{\dagger} \boldsymbol{a}_{\mu} - \boldsymbol{a}_{\mu}^{\dagger} \boldsymbol{a}_{\nu}, \right), \\
        \boldsymbol{L}_{a_{\mu \nu}}^z &= \frac{1}{2} \left( \boldsymbol{a}_{\mu}^{\dagger} \boldsymbol{a}_{\mu} - \boldsymbol{a}_{\nu}^{\dagger} \boldsymbol{a}_{\nu}, \right),
    \end{split}
    \label{eq:SpinOperators}
\end{equation}
acting on the pair of modes $(\boldsymbol{a}_{\mu},\boldsymbol{a}_{\nu})$.
The Casimir operator commuting with all three spin operators is given by the total spin $(\boldsymbol{L}_{a_{\mu \nu}})^2 = (\boldsymbol{L}_{a_{\mu \nu}}^x)^2 + (\boldsymbol{L}_{a_{\mu \nu}}^y)^2 + (\boldsymbol{L}_{a_{\mu \nu}}^z)^2$ and can also be expressed as $(\boldsymbol{L}_{a_{\mu \nu}})^2 = \boldsymbol{L}^0_{a_{\mu \nu}} (\boldsymbol{L}^0_{a_{\mu \nu}} + \mathds{1})$, where the zeroth spin component, defined as
\begin{equation}
    \boldsymbol{L}^0_{a_{\mu \nu}} = \frac{1}{2} \left( \boldsymbol{a}_{\mu}^{\dagger} \boldsymbol{a}_{\mu} + \boldsymbol{a}_{\nu}^{\dagger} \boldsymbol{a}_{\nu} \right),
    \label{eq:L0SpinOperator}
\end{equation}
denotes (one half) the total photon number on the two modes of index $\mu$ and $\nu$.

The zeroth and $z$-components can be measured via photon number measurements as $\braket{\braket{\boldsymbol{L}_{a_{\mu \nu}}^0}} = ( \braket{\boldsymbol{n}_{a_\mu}} + \braket{\boldsymbol{n}_{a_\nu}} )/2$ and $\braket{\braket{\boldsymbol{L}_{a_{\mu \nu}}^z}} = ( \braket{\boldsymbol{n}_{a_\mu}} - \braket{\boldsymbol{n}_{a_\nu}} )/2$, where $\boldsymbol{n}_{a_{\mu}}$ (or $\boldsymbol{n}_{a_\nu}$) denotes the particle number operator associated with mode $\boldsymbol{a}_{\mu}$ (or $\boldsymbol{a}_{\nu}$). Note also that the zeroth component can be measured simultaneously with any other spin operator and will always amount to measuring the total particle number. For the $x$- and $y$-components, simple optical circuits for transforming them into the $z$-component are described in \cite{Hertz2019b,Griffet2022a}, which will be discussed below. We may of course analogously define the spin components $\boldsymbol{L}_{b_{\mu \nu}}^x$, $\boldsymbol{L}_{b_{\mu \nu}}^y$, $\boldsymbol{L}_{b_{\mu \nu}}^z$, and $\boldsymbol{L}_{b_{\mu \nu}}^0$ for any two modes $\boldsymbol{b}_{\mu}$ and $\boldsymbol{b}_{\nu}$ of subsystem $B$, which will be needed in \autoref{subsec:GaussianStates} and \autoref{subsec:MixedSchrödingerCatStates}. We may even define such spin operators across the two subsystems. For a single copy, this amounts to replacing $\boldsymbol{a}_{\mu}$ with $\boldsymbol{a}$ and $\boldsymbol{a}_{\nu}$ with $\boldsymbol{b}$ in Eq. \eqref{eq:SpinOperators}, as we will need in \autoref{subsec:NOONStates}.

\section{Multimode entanglement witnesses}
\label{sec:EntanglementCriteria}
Now we are ready to develop multicopy implementations of the separability criteria from the Shchukin-Vogel hierarchy. In a nutshell, our overall strategy is to identify physically relevant separability criteria from \eqref{eq:ShchukinVogelCriteriaDeterminants}, rewrite them in terms of multimode observables, and then apply linear optical circuits transforming them into spin operators \eqref{eq:SpinOperators}, which can be accessed by photon number measurements following \cite{Hertz2019b,Griffet2022a}. Below, we provide the resulting measurement routines for three classes of criteria that witness entanglement in Gaussian (\autoref{subsec:GaussianStates}), mixed Schrödinger cat (\autoref{subsec:MixedSchrödingerCatStates}), and NOON states (\autoref{subsec:NOONStates}). In each case, we address two potential sources of experimental imperfections, namely imperfect copies and optical losses.

First, remark that multiple identical copies of the state are always assumed to be prepared in the multicopy method. In practice, however, the preparation process encompasses slight fluctuations, so that the prepared multicopy state will contain \textit{imperfect} copies. Although our separability criteria are not guaranteed to remain necessarily valid from first principles in this case, we analyze whether this effect might lead to false-positive detection of entanglement, \textit{i. e.}, might result in a negative determinant even if the imperfect copies are separable. To that end, we model the imperfect preparation by assuming a fixed form of the state -- for instance a Gaussian form -- for all copies but allow the parameters describing the state to differ from copy to copy. Under this assumption, we do not observe any false-positive detections for all the criteria that we have studied. Yet, imperfect copies typically weaken the detection capability of these criteria.

Second, it is clear that any optical setup will suffer from unavoidable losses, which may challenge the multicopy method. We model their effect with a pure-loss channel: each mode of interest $\boldsymbol{a}_\mu$ is coupled with the vacuum $\ket{0}$ via a beam splitter of transmittance $\tau_{a_\mu}$ \cite{Weedbrook2012,Serafini2017}. Effectively, this amounts to multiplying each mode operator $\boldsymbol{a}_\mu$ by $\sqrt{\tau_{a_\mu}}$. As expected, it appears that the detection of entanglement is hindered by such optical losses for all the criteria we have studied. In what follows, we quantify precisely the extent to which these two sources of imperfections affect our criteria.

\subsection{Second-order witness based on $\mathbf{D_{1,2,4}}$} 
\label{subsec:GaussianStates}

\subsubsection{Separability criterion}
We start with the subdeterminant obtained from \eqref{eq:ShchukinVogelCriteriaDeterminants} by selecting the rows and columns 1,2, and 4 of $D^{T_2}$, \textit{i. e.},
\begin{equation}
    d_{1,2,4} = \begin{vmatrix}
    1 & \braket{\boldsymbol{a}} & \braket{\boldsymbol{b}^{\dagger}}\\
    \braket{\boldsymbol{a}^{\dagger}} & \braket{\boldsymbol{a}^{\dagger}\boldsymbol a} & \braket{\boldsymbol{a}^{\dagger}\boldsymbol b^\dagger}\\
    \braket{\boldsymbol b} & \braket{\boldsymbol{a b}} & \braket{\boldsymbol{b}^{\dagger} \boldsymbol{b}}
    \end{vmatrix},
    \label{eq:d124}
\end{equation}
corresponding to the operator $\boldsymbol{f} = c_1 + c_2 \boldsymbol{a} + c_3 \boldsymbol{b}$. As the resulting entanglement witness $d_{1,2,4} < 0$ is of second order in the mode operators, let us compare it to other prominent second-order criteria. To that end, we introduce the non-local quadrature operators \cite{Einstein1935}
\begin{equation}
    \boldsymbol{x}_{\pm} = \abs{r} \, \boldsymbol{x}_1 \pm \frac{1}{r} \, \boldsymbol{x}_2, \quad \boldsymbol{p}_{\pm} = \abs{r} \, \boldsymbol{p}_1 \pm \frac{1}{r} \, \boldsymbol{p}_2, 
\end{equation}
with some real $r \neq 0$. For any separable state, the sums of the variances of these operators are constrained by the criterion of Duan, Giedke, Cirac and Zoller \cite{Duan2000}
\begin{equation}
    d_{\text{Duan}} = \sigma^2_{\boldsymbol{x}_{\pm}} + \sigma^2_{\boldsymbol{p}_{\mp}} - \left(r^2 + \frac{1}{r^2} \right) \ge 0,
    \label{eq:DuanCriteria}
\end{equation}
where $\sigma^2_{\boldsymbol{x}} = \braket{\boldsymbol{x}^2} - \braket{\boldsymbol{x}}^2$ denotes the variance of the operator $\boldsymbol{x}$. Interestingly, the optimized (over $r$) version of condition \eqref{eq:DuanCriteria} is implied by the non-negativity of $d_{1,2,4}$ (see \hyperref[app:DuanCriterion]{Appendix} \ref{app:DuanCriterion} for a proof). In fact, the witness $d_{1,2,4} < 0$ is strictly stronger than the criterion of Duan \textit{et al.} for detecting entanglement.

\subsubsection{Application to Gaussian states}

It is well known that the criterion \eqref{eq:DuanCriteria} is a necessary and sufficient condition for separability (after optimization over $r$) in the case of Gaussian states as considered here (when Alice and Bob hold one mode each) \cite{Horodecki2009}. By the latter considerations, the same holds true for the determinant $d_{1,2,4}$. As a particular example, we evaluate this determinant for the archetypal entangled Gaussian state, the two-mode squeezed vacuum state 
\begin{equation}
    \ket{\psi} = \sqrt{1-\lambda^2} \sum_{n=0}^\infty \lambda^n \ket{n,n},
    \label{eq:TMSVState}
\end{equation}
where $\lambda \in (-1,1)$. 
This leads to the expression
\begin{equation}
    d_{1,2,4} = - \frac{\lambda^2}{1-\lambda^2},
    \label{eq:d124TMSV}
\end{equation}
which is indeed negative for any value of the parameter $\lambda\in (-1,1)$. More details can be found in \hyperref[App:valueTMSV]{Appendix} \ref{App:valueTMSV}.

\begin{figure*}[t!]
    \centering    
    \includegraphics[width=0.999\textwidth]{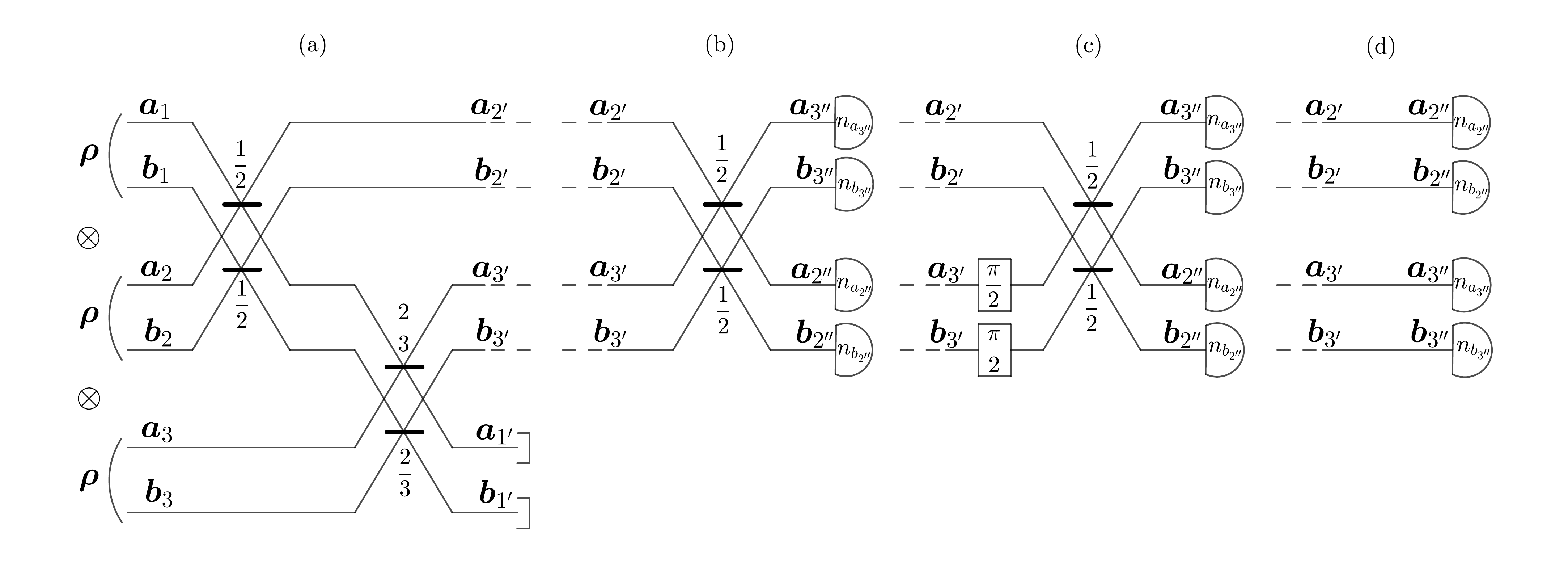}
    \caption{a) Optical circuit implementing the transformation $M$ on three identical copies of the bipartite state $\boldsymbol{\rho}$, where Alice holds modes $\boldsymbol{a}_{1,2,3}$ and Bob holds modes $\boldsymbol{b}_{1,2,3}$. The displacement of the state $\boldsymbol{\rho}$ is removed by a sequence of two beam splitters of transmittances $\frac{1}{2}$ and $\frac{2}{3}$, implemented locally by Alice and Bob, and leading to modes $\boldsymbol{a}_{1',2',3'}$ and $\boldsymbol{b}_{1',2',3'}$.
    The mean field is concentrated on one mode of each subsystem ($\boldsymbol{a}_{1'}$ and $\boldsymbol{b}_{1'}$), which is traced over.         b-d) Three optical circuits  applied locally by Alice and Bob in order to access the expectation values of the three operators $\boldsymbol{C}_j$, which are needed to evaluate the entanglement witness $d_{1,2,4}$. While measuring $\boldsymbol{C}_3$ (see d) requires photon number detectors without any additional optical circuit, a beam splitter of transmittance $\frac{1}{2}$ must be added by Alice and Bob for measuring $\boldsymbol{C}_1$ (see b), preceded by a phase shift of $\frac{\pi}{2}$ for measuring $\boldsymbol{C}_2$ (see c).}
    \label{fig:threecircuits}
\end{figure*}

\subsubsection{Multicopy implementation}
We apply the multicopy measurement method, \textit{i. e.}, assign one copy to each row of the matrix and sum over all permutations, yielding
\begin{equation}
    \boldsymbol D_{1,2,4} = \frac{1}{\abs{S_{123}}}  \sum_{\sigma \in S_{123}}
	\begin{vmatrix}
	1 &  \boldsymbol{a}_{\sigma(1)} &  \boldsymbol{b}^{\dagger}_{\sigma(1)} \\
	\boldsymbol{a}^{\dagger}_{\sigma(2)} &  \boldsymbol{a}^{\dagger}_{\sigma(2)}\boldsymbol a_{\sigma(2)}  &  \boldsymbol{a}^{\dagger}_{\sigma(2)}\boldsymbol b^\dagger_{\sigma(2)} \\
	\boldsymbol b_{\sigma(3)} &  \boldsymbol{a}_{\sigma(3)} \boldsymbol{ b}_ {\sigma(3)}&  \boldsymbol{b}^{\dagger}_{\sigma(3)} \boldsymbol{b}_{\sigma(3)} \\
	\end{vmatrix}, 
	\label{eq:B124}
\end{equation}
where $S_{123}$ denotes the group of permutations over the index set $\{1,2,3\}$ with dimension $\abs{S_{123}} = 3!$. By construction, the multicopy expectation value of this observable gives the determinant \eqref{eq:d124}, \textit{i. e.}, $\braket{\braket{\braket{\boldsymbol D_{1,2,4}}}} = d_{1,2,4}$. 

Since $d_{1,2,4}$ is invariant under displacements (see \hyperref[app:Displacementd124]{Appendix} \ref{app:Displacementd124}), we may access $\boldsymbol{D}_{1,2,4}$ by first applying a linear optics transformation on Alice's and Bob's subsystems that has the effect of concentrating the mean field on one mode of each subsystem ($\boldsymbol{a}_1$ and $\boldsymbol{b}_1$) and canceling it on the other two modes ($\boldsymbol{a}_2$ and $\boldsymbol{a}_3$, on Alice's side, and $\boldsymbol{b}_2$ and $\boldsymbol{b}_3$ on Bob's side). To that end, as shown in Refs. \cite{Hertz2019b,Griffet2022a}, we may 
apply the transformation
\begin{equation}
    \begin{split}
        M & = \Big[ \textrm{BS}_{a_1a_3}\left(2/3\right) \otimes \textrm I_{a_2}\Big] \Big[\textrm{BS}_{a_1a_2}\left(1/2\right) \otimes \textrm I_{a_3} \Big]\\
        & = \frac{1}{\sqrt{6}}\begin{pmatrix}
        \sqrt{2} & \sqrt{2} & \sqrt{2}\\
        \sqrt{3} & -\sqrt{3} & 0\\
        1 & 1 & -2 \\
        \end{pmatrix}
    \end{split}
    \label{eq:transfo-M}
\end{equation}
to the $\boldsymbol{a}$-modes and similarly to the $\boldsymbol{b}$-modes as shown in  \autoref{fig:threecircuits}a. Denoting with a prime all output modes of this transformation, this results in
\begin{equation}
    \begin{split}
        \boldsymbol D_{1,2,4} &= \frac{1}{2} \Big(\boldsymbol{a}^{\dagger}_{2'} \boldsymbol{a}_{2'} \boldsymbol{b}^{\dagger}_{3'} \boldsymbol{b}_{3'} + \boldsymbol{a}^{\dagger}_{3'} \boldsymbol{a}_{3'} \boldsymbol{b}^{\dagger}_{2'} \boldsymbol{b}_{2'} \\
        &\hspace{0.6cm}-\boldsymbol{a}^{\dagger}_{2'} \boldsymbol{a}_{3'} \boldsymbol{b}^{\dagger}_{2'} \boldsymbol{b}_{3'} -\boldsymbol{a}^{\dagger}_{3'} \boldsymbol{a}_{2'} \boldsymbol{b}^{\dagger}_{3'} \boldsymbol{b}_{2'} \Big) \\
        &= \frac{1}{\abs{S_{2'3'}}}  \sum_{\sigma \in S_{2' 3'}}
	       \begin{vmatrix}  \boldsymbol{a}^{\dagger}_{\sigma(1)}\boldsymbol{a}_{\sigma(1)}  &  \boldsymbol{a}^{\dagger}_{\sigma(1)} \boldsymbol{b}^\dagger_{\sigma(1)} \\ \boldsymbol{a}_{\sigma(2)} \boldsymbol{b}_ {\sigma(2)}&  \boldsymbol{b}^{\dagger}_{\sigma(2)} \boldsymbol{b}_{\sigma(2)}
	   \end{vmatrix} \\
        &= \boldsymbol{D}_{2,4},
    \end{split}
    \label{eq:MulticopyAfterRotation}
\end{equation}
with $S_{2' 3'}$ denoting the group of permutations over the index set $\{2',3'\}$ with dimension  $\abs{S_{2' 3'}} = 2!$. Note that the dependence on mode $\boldsymbol{a}_{1'}$ and $\boldsymbol{b}_{1'}$ has disappeared, as expected. Interestingly, the latter expression corresponds to the multicopy implementation of the subdeterminant
\begin{equation}
    d_{2,4} = \begin{vmatrix}
    \braket{\boldsymbol{a}^{\dagger}\boldsymbol a} & \braket{\boldsymbol{a}^{\dagger}\boldsymbol b^\dagger}\\
    \braket{\boldsymbol{a b}} & \braket{\boldsymbol{b}^{\dagger} \boldsymbol{b}}
    \end{vmatrix},
    \label{eq:d24}
\end{equation}
as $\braket{\braket{\boldsymbol{D}_{2,4}}} = d_{2,4}$.

Let us now consider the experimental measurement of the multimode observable $\boldsymbol{D}_{2,4}$.
To that end, we define three operators $\boldsymbol{C}_j$ based on the spin operators \eqref{eq:SpinOperators} and \eqref{eq:L0SpinOperator} applied onto modes $2'$ and $3'$ on Alice's and Bob's side, namely,
\begin{equation}
\label{eq:DefinitionA}
    \begin{split}
        \boldsymbol C_1 &= \boldsymbol{L}_{a_{2'3'}}^0 \boldsymbol{L}_{b_{2'3'}}^0-  \boldsymbol{L}_{a_{2'3'}}^x \boldsymbol{L}_{b_{2'3'}}^x, \\
        \boldsymbol C_2 &= \boldsymbol{L}_{a_{2'3'}}^0 \boldsymbol{L}_{b_{2'3'}}^0-  \boldsymbol{L}_{a_{2'3'}}^y \boldsymbol{L}_{b_{2'3'}}^y, \\
        \boldsymbol C_3 &= \boldsymbol{L}_{a_{2'3'}}^0 \boldsymbol{L}_{b_{2'3'}}^0-  \boldsymbol{L}_{a_{2'3'}}^z \boldsymbol{L}_{b_{2'3'}}^z,
    \end{split}
\end{equation}
leading to the simple decomposition (see \hyperref[app:B24Details]{Appendix} \ref{app:B24Details})
\begin{equation}
    \boldsymbol{D}_{2,4} = \boldsymbol{C}_1 - \boldsymbol{C}_2 + \boldsymbol{C}_3.
    \label{eq:B124UsingA}
\end{equation}
Therefore, $d_{1,2,4}$ can be accessed by measuring separately the expectation value of each operator $\boldsymbol{C}_j$, resulting in
\begin{equation}
    d_{1,2,4} = \braket{\braket{\boldsymbol{C}_1}} - \braket{\braket{\boldsymbol{C}_2}}  + \braket{\braket{\boldsymbol{C}_3}}.
\end{equation}
This is achieved by applying the three linear optical circuits depicted in \autoref{fig:threecircuits}b-d on the  $\boldsymbol{a}$ modes, namely,
\begin{equation}
    \begin{split}
        M_1 &= \textrm{BS}_{a_{2'}a_{3'}}\!\left(1/2\right), \\
        M_2 &= \textrm{BS}_{a_{2'}a_{3'}}\!\left(1/2\right)  \textrm{PS}_{a_{3'}} \!\left(\pi/2\right), \\
        M_3 &= \textrm{I}_{a_{2'}a_{3'}},
    \end{split}
\end{equation}
and analogously for the $\boldsymbol{b}$ modes. Afterwards, all three operators $\boldsymbol{C}_j$ are of the same form 
\begin{equation}    
    \boldsymbol{C}_j = \frac{1}{2} \left( \boldsymbol n_{a_{2''}} \boldsymbol n_{b_{3''}} + \boldsymbol n_{a_{3''}} \boldsymbol n_{b_{2''}} \right),
    \label{eq:AkAfterTransformation}
\end{equation}
which is positive semi-definite and only contains photon number operators (the double primes denote the output modes of the $M_j$ transformations). Thus, the resulting observable  $\boldsymbol{D}_{2,4}$ (hence also $\boldsymbol{D}_{1,2,4}$) depends on cross correlations between the particle numbers on two modes on Alice's and Bob's sides, so it can easily be accessed (provided we have detectors with photon number resolution).

\subsubsection{Imperfect copies and optical losses}
We analyze the influence of imperfect copies and optical losses when applying this witness to the two-mode squeezed vacuum state \eqref{eq:TMSVState}. To that end, we allow for distinct squeezing parameters $\lambda_{\mu} \in (-1,1)$ for the two copies $\mu=1,2$. Thus, we consider the state $\ket{\psi}\bra{\psi}_1 \otimes \ket{\psi}\bra{\psi}_2$ and insert  beam splitters with transmittances $\tau_{a_{\mu}}, \tau_{b_{\mu}} \leq 1$ on the four modes $(\boldsymbol{a}_1,\boldsymbol{a}_2)$ and $(\boldsymbol{b}_1,\boldsymbol{b}_2)$ in order to model losses. Then, we obtain for the expectation value of our multicopy observable $\boldsymbol{D}_{2,4}$
\begin{equation}
    \begin{split}
\braket{\braket{\boldsymbol{D}_{2,4}}} &= \frac{\lambda_1^2 \lambda_2^2 (\tau_{a_1} \tau_{b_2} + \tau_{a_2} \tau_{b_1}) - 2\lambda_1 \lambda_2 \sqrt{\tau_{a_1} \tau_{a_2}  \tau_{b_1} \tau_{b_2}}}{2(1-\lambda_1^2)(1-\lambda_2^2)},
     \end{split}
\end{equation}
with a slight abuse of notation (we use double brackets although the two copies are not identical). We note first that, without losses, the multicopy expectation value 
\begin{equation}
    \begin{split}
\braket{\braket{\boldsymbol{D}_{2,4}}}^\text{no-loss} &= \frac{\lambda_1 \lambda_2 (\lambda_1 \lambda_2 - 1)}{(1-\lambda_1^2)(1-\lambda_2^2)} 
     \end{split}
\end{equation}
is always negative provided $\lambda_1$ and $\lambda_2$ have the same sign. Otherwise, if $\lambda_1$ and $\lambda_2$ have opposite signs, entanglement is not detected anymore (this corresponds to false negatives, \textit{i. e.}, the determinant fails to be negative even if the imperfect copies are both entangled). We also see that  $\braket{\braket{\boldsymbol{D}_{2,4}}}^\text{no-loss} = 0$ if  $\lambda_1 = 0$ or $\lambda_2 = 0$, in which case the state $\ket{\psi}_1$ or $\ket{\psi}_2$ becomes trivially separable and hence, we do not get a false-positive detection of entanglement.

Now adding losses but assuming that $\tau_{a_1} = \tau_{a_2} = \tau_{b_1} = \tau_{b_2} \equiv \tau$, we get the expectation value 
\begin{equation}
    \begin{split}
\braket{\braket{\boldsymbol{D}_{2,4}}}
 = \tau^2 \braket{\braket{\boldsymbol{D}_{2,4}}}^\text{no-loss}  .
     \end{split}
\end{equation}
Thus, in the interesting case where $\lambda_1$ and $\lambda_2$ have the same sign, the no-loss negative value is multiplied by a positive factor $\tau^2$ smaller than or equal to unity.
This implies that losses can only deteriorate the detection capabilities but, at the same time, the two-mode squeezed vacuum state remains detected for any non-vanishing transmittances $\tau > 0$. More generally, using $(\sqrt{\tau_{a_1} \tau_{b_2}} \pm \sqrt{\tau_{a_2} \tau_{b_1}})^2\geq 0$, we obtain upper and lower bounds on the expectation value of $\boldsymbol{D}_{2,4}$ with arbitrary  losses, namely
\begin{equation}
    \begin{split}
        &\frac{\tau_{a_1} \tau_{b_2} + \tau_{a_2} \tau_{b_1}}{2} ~ \braket{\braket{\boldsymbol{D}_{2,4}}}^\text{no-loss}  \ge \braket{\braket{\boldsymbol{D}_{2,4}}} \\
        & ~~~~~~~~~~~~~~~ \ge \sqrt{\tau_{a_1} \tau_{a_2} \tau_{b_1} \tau_{b_2}} ~ \braket{\braket{\boldsymbol{D}_{2,4}}}^\text{no-loss} ,
    \end{split}
    \label{eq:D24Bounds}
\end{equation}
\begin{figure*}[t!]
    \centering    
    \includegraphics[width=0.999\textwidth]{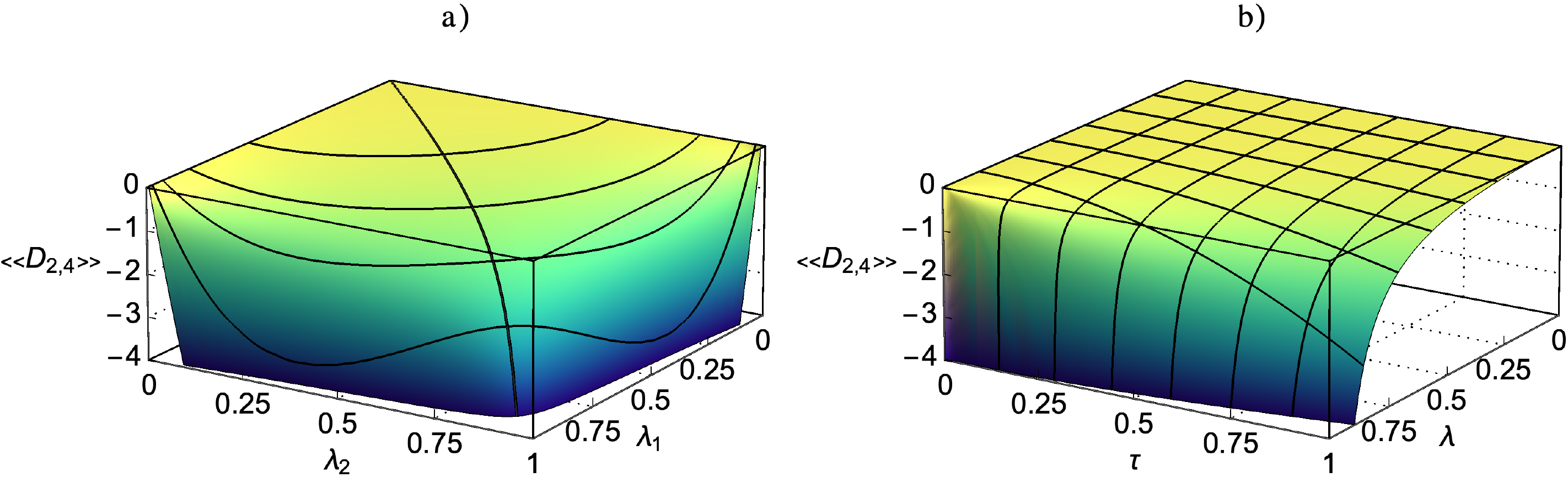}
    \caption{a) Expectation value of the multicopy observable $\boldsymbol{D}_{2,4}$ as a function of the squeezing parameters $\lambda_1, \lambda_2$ describing two different two-mode squeezed vacuum states, with contour lines of equal total entanglement entropy. The diagonal line corresponds to $\lambda_1=\lambda_2$. We observe that $\braket{\braket{\boldsymbol{D}_{2,4}}}$ is negative for all $\lambda_1, \lambda_2 > 0$, but identical copies do not minimize $\braket{\braket{\boldsymbol{D}_{2,4}}}$ for a given amount of entanglement. b) Dependence of $\braket{\braket{\boldsymbol{D}_{2,4}}}$ on losses for given squeezing $\lambda$. As expected, decreasing the transmittance $\tau$ makes the value of $\braket{\braket{\boldsymbol{D}_{2,4}}}$ approach zero, but it remains negative for all $\tau > 0$.}
    \label{fig:D24Experiment}
\end{figure*}
where we have assumed again that $\lambda_1$ and $\lambda_2$ have the same sign, so that $\braket{\braket{\boldsymbol{D}_{2,4}}}^\text{no-loss}$ is negative. Both bounds simply collapse to $\tau^2 \braket{\braket{\boldsymbol{D}_{2,4}}}^\text{no-loss}$ in the case where $\tau_{a_1} \tau_{b_2}= \tau_{a_2} \tau_{b_1} \equiv \tau^2$, from which we draw the same conclusions. Otherwise, for arbitrary transmittances, it is clear that losses always bring the (negative) lower bound on $\braket{\braket{\boldsymbol{D}_{2,4}}}$ closer to zero, corroborating the idea that losses deteriorate the witness. Yet, the two-mode squeezed vacuum state remains detected for any non-vanishing transmittance $\tau_{a_\mu}, \tau_{b_{\mu}} > 0$ as the upper bound on $\braket{\braket{\boldsymbol{D}_{2,4}}}$ always remains negative. In short, although the condition $\braket{\braket{\boldsymbol{D}_{2,4}}}^\text{no-loss} < 0$ only constitutes a valid entanglement witness for $\lambda_1 = \lambda_2$, we observe that $\braket{\braket{\boldsymbol{D}_{2,4}}}$ is negative for all $\lambda_1, \lambda_2 > 0$ or $\lambda_1, \lambda_2 < 0$ and for arbitrary losses. 

We can further illustrate the fact that the false-positive detection of entanglement is excluded by considering a finite set of separable states. For example, for two imperfect copies of a product of two single-mode squeezed states, the expectation value is given by
\begin{equation}
    \begin{split}
        \braket{\braket{\boldsymbol{D}_{2,4}}}  &= \frac{1}{2} \Big(\tau_{a_1} \tau_{b_2} \sinh^2 r_{a_1} \sinh^2 r_{b_2}  \\
        & ~~~~~ + \tau_{b_1} \tau_{a_2} \sinh^2 r_{a_2} \sinh^2 r_{b_1} \Big),
    \end{split}    
\end{equation}
where $r_{a_\mu}, \ r_{b_\mu} \in [0,\infty)$ are the squeezing parameters of the four single-mode squeezed states injected in the circuit. This expression is always non-negative and hence, we cannot obtain a false-positive detection of entanglement. 

We have plotted the dependence of the muticopy expectation value $\braket{\braket{\boldsymbol{D}_{2,4}}}$ on the two squeezing parameters $\lambda_1$ and $\lambda_2$ in the no-loss case in \autoref{fig:D24Experiment}a, together with contour lines of equal total entanglement entropy and a diagonal line along $\lambda_1 = \lambda_2$ indicating identical copies. Although false-positive detection is excluded, we observe that the observable is \textit{not} jointly convex in $\lambda_1$ and $\lambda_2$ for a fixed amount of entanglement as the case $\lambda_1 = \lambda_2$ corresponds to a local maximum (instead of a global minimum) along every contour line of fixed total entanglement entropy. However, since the non-negativity of $\braket{\braket{\boldsymbol{D}_{2,4}}}$ does only constitute a separability criterion if the two copies are perfect, the possibility that its value decreases (becomes more negative) for imperfect copies is acceptable as long as both states remain entangled.

The effect of losses is illustrated for $\tau \equiv \tau_{a_1} = \tau_{a_2} = \tau_{b_1} = \tau_{b_2}$ and perfect copies $\lambda \equiv \lambda_1 = \lambda_2$ in \autoref{fig:D24Experiment}b, together with contours of equal $\tau$ and equal $\lambda$. For decreasing $\tau$, the value of $\braket{\braket{\boldsymbol{D}_{2,4}}}$ falls off quadratically and attains zero for $\tau = 0$, \textit{i. e.}, when the input signal is fully lost. This detrimental effect of losses is clearly stronger when the state is more entangled.

\subsection{Fourth-order witness based on $\mathbf{D_{1,4,9}}$}
\label{subsec:MixedSchrödingerCatStates}

\subsubsection{Separability criterion}

We now consider the criterion obtained from the operator $\boldsymbol{f} = c_1 + c_2 \boldsymbol{b} + c_3 \boldsymbol{a} \boldsymbol{b}$, corresponding to the determinant (see \cite{Shchukin2005} for the ordering convention of moments)
\begin{equation}
    d_{1,4,9} = \begin{vmatrix}
    1 & \braket{\boldsymbol{b}^{\dagger}} & \braket{\boldsymbol{a} \boldsymbol{b}^{\dagger}}\\
    \braket{\boldsymbol{b}} & \braket{\boldsymbol{b}^{\dagger} \boldsymbol{b}} & \braket{\boldsymbol{a} \boldsymbol{b}^{\dagger} \boldsymbol{b}}\\
    \braket{\boldsymbol{a}^{\dagger} \boldsymbol{b}} & \braket{\boldsymbol{a}^{\dagger} \boldsymbol{b}^{\dagger} \boldsymbol{b}} & \braket{\boldsymbol{a}^{\dagger} \boldsymbol{a} \boldsymbol{b}^{\dagger} \boldsymbol{b}}
    \end{vmatrix}  .
    \label{eq:d149}
\end{equation}
This determinant is of fourth order in the mode operators and thus the corresponding witness $d_{1,4,9} < 0$ is of particular interest for detecting non-Gaussian entanglement.

\subsubsection{Application to mixed Schrödinger cat states}
We introduce the general family of two-mode Schrödinger cat states obtained by superposing two pairs of coherent states $\ket{\alpha, \beta}$ and $\ket{-\alpha, -\beta}$, namely 
\begin{equation}
    \begin{split}
        \boldsymbol{\rho} &= N (\alpha, \beta, z) \Big[ \ket{\alpha, \beta} \bra{\alpha, \beta} + \ket{-\alpha, -\beta} \bra{-\alpha, -\beta} \\
        &- (1-z)  \left( \ket{\alpha, \beta} \bra{-\alpha, -\beta} + \ket{-\alpha, -\beta} \bra{\alpha, \beta} \right) \Big],
    \end{split}
    \label{eq:CatState}
\end{equation}
with a mixing parameter $z \in [0,1]$ and a normalization constant $N (\alpha, \beta, z) = \left(1-(1-z) e^{-2 (\abs{\alpha^2} + \abs{\beta}^2)} \right)^{-1}/2$, with $\alpha, \beta \in \mathbb{C}$. The state \eqref{eq:CatState} is pure if and only if $z=0$, in which case it reduces to the cat state considered in Ref. \cite{Shchukin2005}, while for $z>0$ it corresponds to a mixed cat state. The special case $\alpha = \beta$ has been considered in Refs. \cite{Walborn2009,Saboia2011,Haas2022a}. Further, state \eqref{eq:CatState} is separable if and only if $z=1$ or $\alpha = \beta = 0$ (in which case it corresponds to the vacuum provided $z \neq 0$; it is ill-defined for $z = 0$ in this case). While second-moment criteria can not certify entanglement at all, sophisticated entropic criteria witness entanglement only for sufficiently large $\abs{\alpha} = \abs{\beta} \gtrsim 3/2$ \cite{Walborn2009,Saboia2011,Haas2022d}, in which case \eqref{eq:CatState} corresponds to two well separated coherent states. 

In contrast, the determinant \eqref{eq:d149} evaluates to (details of this calculation can be found in \hyperref[App:valueCat]{Appendix} \ref{App:valueCat})
\begin{equation} 
    d_{1,4,9} = - \abs{\alpha}^2 \abs{\beta}^4 \frac{\coth \left[ \abs{\alpha}^2 + \abs{\beta}^2 - \frac{1}{2} \ln \left(1-z \right) \right]}{\sinh^2 \left[ \abs{\alpha}^2 + \abs{\beta}^2 - \frac{1}{2} \ln \left(1-z \right) \right]}.
    \label{eq:d149cat}
\end{equation}
As hyperbolic functions map positive numbers to positive numbers, entanglement is certified for the full parameter range, \textit{i. e.} all $z \in [0,1)$ and $\alpha, \beta \in \mathbb{C} \, \backslash \{0\}$. However, this criterion deteriorates when $\alpha$ and $\beta$ become separated due to the limited order of the moments involved in the criterion. Otherwise, for $\alpha$ close to $\beta$, the witness $d_{1,4,9} < 0$ outperforms all known entropic witnesses in the case of cat-like entanglement.

\begin{figure*}[t!]
    \centering    
    \includegraphics[width=0.999\textwidth]{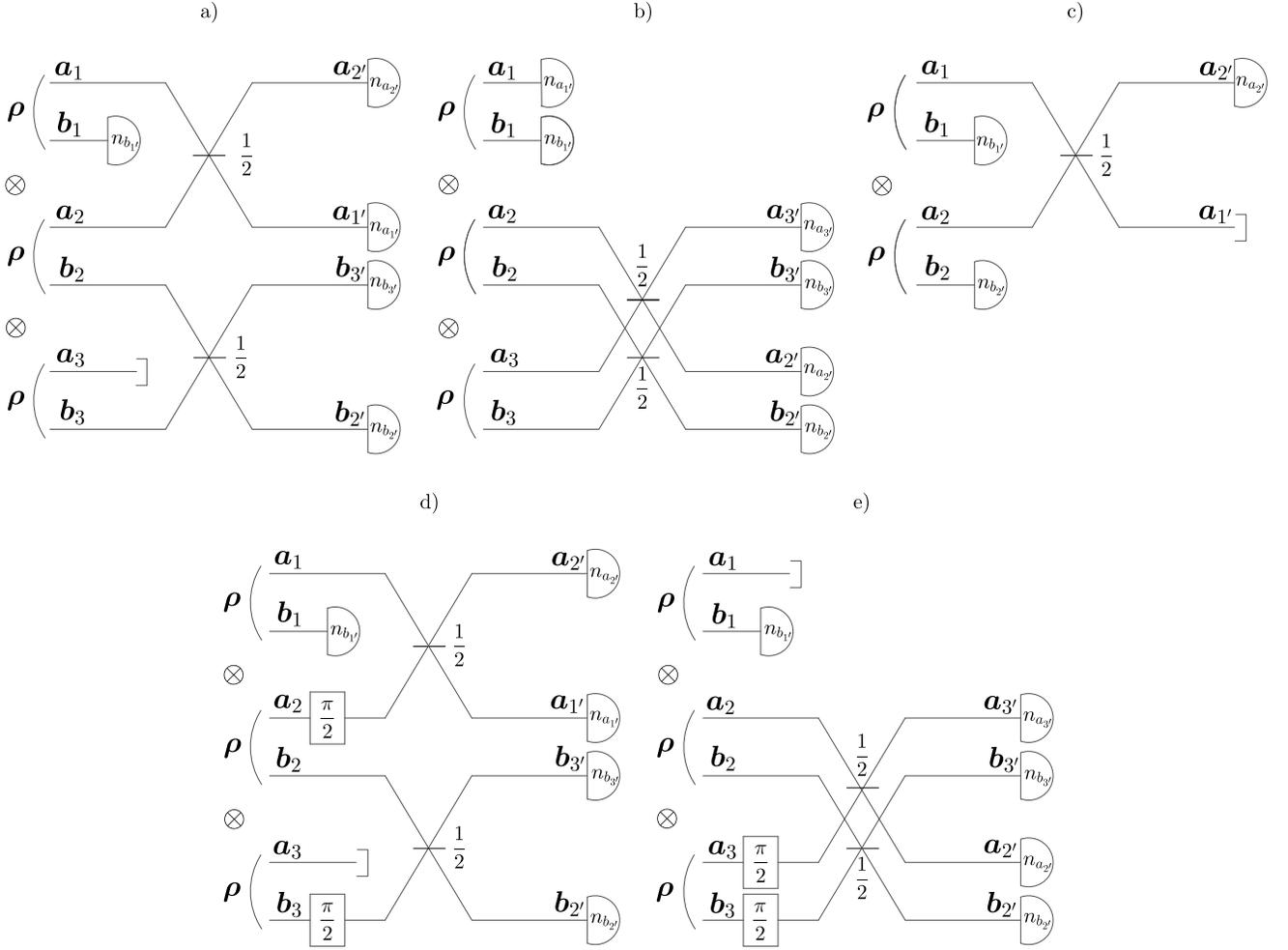}
    \caption{Optical circuits implementing the five transformations $M_j$ that are needed for translating the measurements of the five multicopy observables $\boldsymbol {F}_j$ into photon number measurements. The expectation value $\braket{\braket{\braket{\boldsymbol {F}_j}}}$ is accessed after applying $M_j$ from a) to e), respectively. In all cases, Alice and Bob must apply local transformations to their respective subsystems (Alice holds modes $\boldsymbol{a}_{1,2,3}$ and Bob holds modes $\boldsymbol{b}_{1,2,3}$). Note that the third copy is not needed for the measurement of $\braket{\braket{\braket{\boldsymbol {F}_3}}}$ (see c). }
    \label{fig:d149}
\end{figure*}

\subsubsection{Multicopy implementation}
To efficiently access $d_{1,4,9}$, we again exploit the multicopy method and define the corresponding multicopy observable as
\begin{equation}
    \begin{split}
        &\boldsymbol D_{1,4,9} = \frac{1}{\abs{S_{123}}} \sum_{\sigma \in S_{123}} \\
        &\begin{vmatrix}
        1 & \boldsymbol{b}^{\dagger}_{\sigma(1)} & \boldsymbol{a}_{\sigma(1)} \boldsymbol{b}^{\dagger}_{\sigma(1)}\\
        \boldsymbol{b}_{\sigma(2)} & \boldsymbol{b}^{\dagger}_{\sigma(2)} \boldsymbol{b}_{\sigma(2)} & \boldsymbol{a}_{\sigma(2)} \boldsymbol{b}^{\dagger}_{\sigma(2)} \boldsymbol{b}_{\sigma(2)}\\
        \boldsymbol{a}^{\dagger}_{\sigma(3)} \boldsymbol{b}_{\sigma(3)} & \boldsymbol{a}^{\dagger}_{\sigma(3)} \boldsymbol{b}^{\dagger}_{\sigma(3)} \boldsymbol{b}_{\sigma(3)} & \boldsymbol{a}^{\dagger}_{\sigma(3)} \boldsymbol{a}_{\sigma(3)} \boldsymbol{b}^{\dagger}_{\sigma(3)} \boldsymbol{b}_{\sigma(3)}
        \end{vmatrix},
    \end{split}
	\label{eq:B149}
\end{equation}
such that $d_{1,4,9} = \braket{\braket{\braket{\boldsymbol D_{1,4,9}}}}$. Equation \eqref{eq:B149} consists of 36 terms and can be rewritten as
\begin{equation}
    \boldsymbol{D}_{1,4,9} =\boldsymbol {F}_1 -\boldsymbol {F}_2 + \boldsymbol {F}_3 - \boldsymbol {F}_4 - \boldsymbol {F}_5,
\end{equation}
after defining the five operators
\begin{equation}
    \begin{split}
        \boldsymbol F_1 &=  \frac{1}{\abs{P_{123}}}\sum_{\sigma \in P_{123}} \Big( \boldsymbol L^x_{a_{\sigma(1) \sigma(2)}} + \boldsymbol L^x_{a_{\sigma(3) \sigma(1)}}\Big) \boldsymbol n_{{b_\sigma(1)}}\boldsymbol L^x_{b_{\sigma(2) \sigma(3)}}, \\
        \boldsymbol F_2 &=  \frac{1}{\abs{P_{123}}}\sum_{\sigma \in P_{123}} \Big( \boldsymbol L^x_{a_{\sigma(2) \sigma(3)}} + \boldsymbol n_{a_{\sigma(1)}} \Big) \boldsymbol n_{{b_\sigma(1)}}\boldsymbol L^x_{b_{\sigma(2) \sigma(3)}},\\
        \boldsymbol F_3 &= \frac{1}{\abs{P_{123}}}\sum_{\sigma \in P_{123}} \left( \boldsymbol L^0_{a_{\sigma(1) \sigma(2)}} -\boldsymbol  L^x_{a_{\sigma(1) \sigma(2)}}\right) \boldsymbol n_{b_{\sigma(1)}}\boldsymbol n_{b_{\sigma(2)}}, \\
        \boldsymbol F_4 &= \frac{1}{\abs{P_{123}}}\sum_{\sigma \in P_{123}} \Big( \boldsymbol L^y_{a_{\sigma(1) \sigma(2)}} + \boldsymbol L^y_{a_{\sigma(3) \sigma(1)}} \Big) \boldsymbol n_{b_{\sigma(1)}} \boldsymbol L^y_{b_{\sigma(2)\sigma(3)}},\\
        \boldsymbol F_5 &= \frac{1}{\abs{P_{123}}}\sum_{\sigma \in P_{123}} \boldsymbol L^y _{a_{\sigma(2) \sigma(3)}} \boldsymbol n_{b_{\sigma(1)}} \boldsymbol L^y_{b_{\sigma(2)\sigma(3)}},\\
    \end{split}
\end{equation}
where $P_{123}$ denotes the group of \textit{cyclic} permutations over the index set $\{1,2,3\}$. Thus, the determinant $d_{1,4,9}$ can be accessed by measuring separately the expectation value of each of the  five operators $\boldsymbol {F}_j$, that is,
\begin{equation}
    \begin{split}
        d_{1,4,9} =&  \braket{\braket{\braket{\boldsymbol {F}_1}}} - \braket{\braket{\braket{\boldsymbol {F}_2}}} + \braket{\braket{\braket{\boldsymbol {F}_3}}} \\
        &- \braket{\braket{\braket{\boldsymbol {F}_4}}} - \braket{\braket{\braket{\boldsymbol {F}_5}}} .
    \end{split}
\end{equation}
Fortunately, the multicopy expectation values $\braket{\braket{\braket{\boldsymbol F_j}}}$ simplify by using the symmetry under permutations for the three summands in every operator $\boldsymbol F_j$ as well as for the spin operators themselves. This  leads to
\begin{equation}
    \begin{split}
        \braket{\braket{\braket{\boldsymbol F_1}}} &= 2 \braket{\braket{\braket{\boldsymbol L^x_{a_{12}} \boldsymbol n_{b_{1}} \boldsymbol L^x_{b_{23}}}}}, \\
        \braket{\braket{\braket{\boldsymbol F_2}}} &=  \braket{\braket{\braket{\left( \boldsymbol L^x_{a_{23}} + \boldsymbol n_{a_{1}}\right)\boldsymbol n_{{b_1}}\boldsymbol L^x_{b_{23}}}}}, \\
        \braket{\braket{\braket{\boldsymbol F_3}}} &=  \braket{\braket{\braket{\left( \boldsymbol L^0_{a_{12}} -\boldsymbol  L^x_{a_{12}}\right) \boldsymbol n_{b_{1}}\boldsymbol n_{b_{2}}}}}, \\
        \braket{\braket{\braket{\boldsymbol F_4}}} &= 2 \braket{\braket{\braket{\boldsymbol L^y_{a_{12}}\boldsymbol n_{b_{1}} \boldsymbol L^y_{b_{23}}}}}, \\
        \braket{\braket{\braket{\boldsymbol F_5}}} &=  \braket{\braket{\braket{\boldsymbol L^y_{a_{23}}   \boldsymbol n_{b_{1}} \boldsymbol L^y_{b_{23}}}}}.
    \end{split}
\end{equation}
These five multicopy expectation values can be expressed in terms of photon number measurements by applying the five respective transformations shown in \autoref{fig:d149}a-e, namely
\begin{equation}
    \begin{split}
        M_1 &= \textrm{BS}_{a_1a_2}\left(1/2\right)\textrm{BS}_{b_2b_3}\left(1/2\right)\otimes I_{b_1a_3}, \\
        M_2 &= \textrm{BS}_{a_2a_3}\left(1/2\right)\textrm{BS}_{b_2b_3}\left(1/2\right)\otimes I_{a_1b_1}, \\
        M_3 &= \textrm{BS}_{a_1a_2}\left(1/2\right)\otimes I_{b_1b_2a_3b_3},\\
        M_4 &=
        \textrm{BS}_{a_1a_2}\left(1/2\right) \textrm{BS}_{b_2b_3}\left(1/2\right) \textrm{PS}_{a_2} \left(\pi/2\right)\\
        &\hspace{0.8cm}\textrm{PS}_{b_3} \left(\pi/2\right)\otimes I_{b_1a_3}, \\
        M_5 &= \textrm{BS}_{a_2a_3}\left(1/2\right) \textrm{BS}_{b_2b_3}\left(1/2\right)\textrm{PS}_{a_3} \left(\pi/2\right)\\
        &\hspace{0.8cm} \textrm{PS}_{b_3} \left(\pi/2\right)\otimes I_{a_1b_1}  .
    \end{split}
\end{equation}
Incidentally, we note that the measurement of $\boldsymbol {F}_3$, implemented via $M_3$ (see \autoref{fig:d149} c), only requires two copies, while the other four multicopy observables $\boldsymbol F_j$ are read out on three copies. Then, we finally obtain
\begin{equation}
    \begin{split}
        \braket{\braket{\braket{\boldsymbol F_1}}} &= \frac{1}{2} \braket{\left( \boldsymbol n_{a_{1'}}-\boldsymbol n_{a_{2'}} \right)\boldsymbol n_{{b_{1'}}}\left( \boldsymbol n_{b_{2'}}-\boldsymbol n_{b_{3'}} \right)}, \\
        \braket{\braket{\braket{\boldsymbol F_2}}} &= \frac{1}{2} \langle\Big(\frac{1}{2} \big(\boldsymbol n_{a_{2'}}-\boldsymbol n_{a_{3'}} \big) +\boldsymbol n_{{a_{1'}}}\Big)\\
        &\hspace{0.5cm}\boldsymbol n_{{b_{1'}}}\left( \boldsymbol n_{b_{2'}}-\boldsymbol n_{b_{3'}} \right)\rangle, \\
        \braket{\braket{\braket{\boldsymbol F_3}}} &= \braket{\boldsymbol n_{a_{2'}}\boldsymbol n_{b_{1'}}\boldsymbol n_{b_{2'}}}, \\
        \braket{\braket{\braket{\boldsymbol F_4}}} &= \frac{1}{2} \braket{\left( \boldsymbol n_{a_{1'}}-\boldsymbol n_{a_{2'}} \right)\boldsymbol n_{b_{1'}} \left( \boldsymbol n_{b_{2'}}-\boldsymbol n_{b_{3'}} \right)},\\
        \braket{\braket{\braket{\boldsymbol F_5}}} &= \frac{1}{4} \braket{\left( \boldsymbol n_{a_{2'}}-\boldsymbol n_{a_{3'}} \right)\boldsymbol n_{b_{1'}}\left( \boldsymbol n_{b_{2'}}-\boldsymbol n_{b_{3'}} \right)}.
    \end{split}
\end{equation}

\subsubsection{Imperfect copies and optical losses}
\begin{figure*}[t!]
    \centering    
    \includegraphics[width=0.999\textwidth]{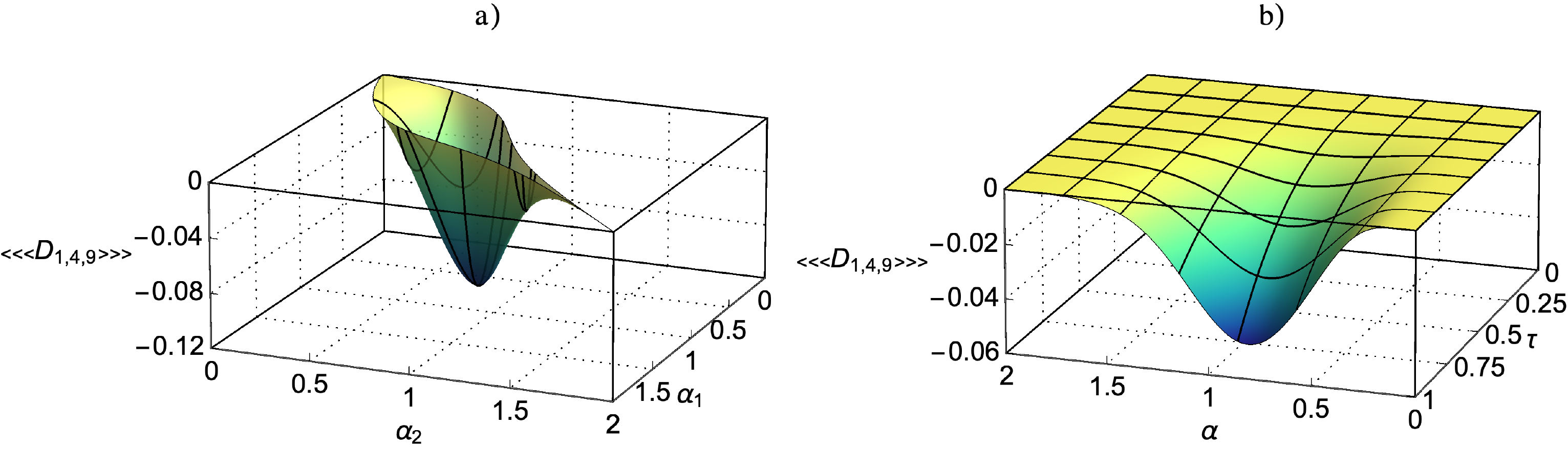}
    \caption{a) Negative regions of the expectation value of the multicopy observable $\boldsymbol{D}_{1,4,9}$ for the mixed Schrödinger cat state with real $\alpha_{\mu} = \beta_{\mu}$, $z_{\mu} = 1/2$, for $\mu=1,2,3$, and unequal first and second copies  ($\alpha_1\ne \alpha_2$) while the third copy is equal to the second one ($\alpha_3 = \alpha_2$). The expectation value $\braket{\braket{\braket{\boldsymbol{D}_{1,4,9}}}}$ remains negative only in a small region around $\alpha_1 \approx \alpha_2$. b) Multicopy expectation value $\braket{\braket{\braket{\boldsymbol{D}_{1,4,9}}}}$ as a function of transmittance $\tau$ and amplitude $\alpha$. Entanglement detection works best around $\alpha \approx 1$ and increasing losses also increase the (negative) value of  $\braket{\braket{\braket{\boldsymbol{D}_{1,4,9}}}}$ without breaking its negativity.}
    \label{fig:D149Experiment}
\end{figure*}

The general expression for $d_{1,4,9}$ when considering three distinct copies and including losses can be found in \hyperref[app:d149imperfections]{Appendix} \ref{app:d149imperfections}.
Here, we restrict our analysis to the special case where all states are equally mixed $z_{\mu}=1/2$ and comprise equal pairs of real amplitudes $\alpha_\mu \equiv \beta_\mu \in \mathbb{R}$, and where all modes undergo equal losses $\tau \equiv \tau_{a_\mu} = \tau_{b_\mu}$, for $\mu=1,2,3$. We analyze the behavior of $\braket{\braket{\braket{\boldsymbol{D}_{1,4,9}}}}$ for \textit{two} different input states (we take copies $2$ and $3$ to be equal but distinct from copy $1$) without losses in \autoref{fig:D149Experiment}a. We observe that if $\alpha_1$ and $\alpha_2$ are too distinct, $\braket{\braket{\braket{\boldsymbol{D}_{1,4,9}}}}$ becomes positive, hence entanglement is undetected. This sensitivity to $|\alpha_1- \alpha_2|$ is very strong for $\alpha_{1,2} \gtrsim 3/2$. Yet, false-positive detection is excluded since $\braket{\braket{\braket{\boldsymbol{D}_{1,4,9}}}} \ge 0$ if $\alpha_1 = 0$ or $\alpha_2 = 0$. 

The case of perfect copies with equal losses in all modes is considered in \autoref{fig:D149Experiment}b, where we plot the dependence of $\braket{\braket{\braket{\boldsymbol{D}_{1,4,9}}}}$ on the transmittance $\tau$ for a given $\alpha$. As expected, losses make the value of $\braket{\braket{\braket{\boldsymbol{D}_{1,4,9}}}}$ approach zero from below for all $\alpha$, but $\braket{\braket{\braket{\boldsymbol{D}_{1,4,9}}}}$ remains negative for all $\tau > 0$ (of course, we have $\braket{\braket{\braket{\boldsymbol{D}_{1,4,9}}}} = 0$ for $\tau = 0$). Also, we observe that the witness $d_{1,4,9}<0$  works best around $\alpha \approx 1$, \textit{i. e.}, if the two coherent states partially overlap. Note here that $\braket{\braket{\braket{\boldsymbol{D}_{1,4,9}}}}$ approaches zero exponentially (from below) for $\alpha \to \infty$, so that entanglement remains witnessed for all $\alpha > 0$.

\subsection{Fourth-order witness based on $\mathbf{D_{1,9,13}}$}
\label{subsec:NOONStates}

\subsubsection{Separability criterion}
We finally consider the separability criterion corresponding to the operator $\boldsymbol f = c_1 + c_2 \boldsymbol a \boldsymbol b + c_3 \boldsymbol a^\dagger \boldsymbol b^\dagger$, \textit{i.e.}, selecting the rows and columns 1, 9, and 13 in \eqref{eq:ShchukinVogelCriteriaDeterminants}, leading to the determinant  
\begin{equation}\label{eq:d1913}
    d_{1,9,13} = \begin{vmatrix}
    1 & \braket{\boldsymbol a \boldsymbol{b}^{\dagger}} & \braket{\boldsymbol{a}^{\dagger} \boldsymbol{b}}\\
    \braket{\boldsymbol a^\dagger \boldsymbol{b}} & \braket{\boldsymbol{a}^{\dagger} \boldsymbol{a} \boldsymbol{b}^{\dagger} \boldsymbol{b}} & \braket{\boldsymbol{a}^{\dagger 2} \boldsymbol{b}^{2}}\\
    \braket{\boldsymbol{a} \boldsymbol{b}^{\dagger}} & \braket{\boldsymbol{a}^{2} \boldsymbol{b}^{\dagger 2} } & \braket{\boldsymbol{a} \boldsymbol{a}^{\dagger} \boldsymbol{b} \boldsymbol{b}^{\dagger}}
    \end{vmatrix}.
\end{equation}
The resulting entanglement witness $d_{1,9,13} < 0$ is again of fourth order in the mode operators. However, when expanding the determinant $d_{1,9,13}$, several products of fourth-order expectation values appear, which gives an overall expression of higher order when compared to $d_{1,4,9}$ in Eq. \eqref{eq:d149}. As we may anticipate, the corresponding multicopy observable will therefore be quite complex.

\subsubsection{Application to NOON states}
In order to illustrate this entanglement witness, we consider the class of pure NOON states with arbitrary complex amplitudes \cite{Sanders1989}
\begin{equation}
    \ket{\psi} = \alpha \ket{n,0} + \beta \ket{0,n},
    \label{eq:NOONState}
\end{equation}
with integer $n \ge 1$ and  $\abs{\alpha}^2 + \abs{\beta}^2 = 1$. Note that this class includes the first Bell state $n=1, \alpha=\beta=1/\sqrt{2}$, as well as the Hong-Ou-Mandel state $n=2, \alpha = - \beta = 1/\sqrt{2}$. The state \eqref{eq:NOONState} is entangled for all allowed parameter values except when $\alpha$ or $\beta$ is equal to zero. However, entanglement can not be witnessed by any second-order nor entropic criterion that is valid for mixed states. Pure state entropic criteria flag entanglement for low excitations, \textit{i. e.} small $n$, see e.g. \cite{Walborn2009,Saboia2011,Haas2022d}, while the Wehrl mutual information fully detects entanglement as it corresponds to a perfect witness for pure states \cite{Haas2021}. In this sense, detecting the entanglement of the NOON states \eqref{eq:NOONState} is known to be particularly challenging, even for small $n$.

When evaluating the determinant \eqref{eq:d1913} for state \eqref{eq:NOONState}, we find (see \hyperref[App:valueNoon]{Appendix} \ref{App:valueNoon} for details)
\begin{equation}
    d_{1,9,13} = - 2 \, \abs{\alpha}^2 \abs{\beta}^2 \left( \delta_{n 1} + 2 \delta_{n 2} \right).
    \label{eq:d1913Noon}
\end{equation}
Thus, the witness $d_{1,9,13} < 0$ flags entanglement for all NOON states with $\alpha, \beta \in \mathbb{C}$ (except when $\alpha =0$ or $\beta = 0$) when $n=1,2$. 

Unfortunately, the straightforward application of the multicopy method leads to an observable $\boldsymbol D_{1,9,13}$ which cannot be accessed by using linear interferometers and photon number measurements, see \hyperref[app:d1913]{Appendix} \ref{app:d1913}. Therefore, we instead consider the weaker criterion $d'_{1,9,13} \ge d_{1,9,13}$, which has been put forward in Ref. \cite{Agarwal2005} (see also \cite{Shchukin2005}) and relies on the same moments as $d_{1,9,13}$, namely, 
\begin{equation}
    \label{eq:agarwal}
    \begin{split}
        d'_{1,9,13} = \, &\Big( \braket{\boldsymbol a^{\dagger} \boldsymbol a \boldsymbol b^{\dagger} \boldsymbol b} + \braket{\boldsymbol a \boldsymbol a^{\dagger} \boldsymbol b \boldsymbol b^{\dagger}} + \braket{\boldsymbol a^{\dagger 2} \boldsymbol b^2} \\
        &\hspace{0.25cm}+ \braket{\boldsymbol a^2 \boldsymbol b^{\dagger2}} - \braket{\boldsymbol a^{\dagger} \boldsymbol b + \boldsymbol a \boldsymbol b^{\dagger}}^2 \Big) \\
        &\Big( \braket{\boldsymbol a^{\dagger} \boldsymbol a \boldsymbol b^{\dagger} \boldsymbol b} + \braket{\boldsymbol a \boldsymbol a^{\dagger} \boldsymbol b \boldsymbol b^{\dagger}} - \braket{\boldsymbol a^{\dagger 2} \boldsymbol b^2} \\
        &\hspace{0.25cm}- \braket{\boldsymbol a^2 \boldsymbol b^{\dagger2}} + \braket{\boldsymbol a^{\dagger} \boldsymbol b - \boldsymbol a \boldsymbol b^{\dagger}}^2 \Big) \\
        &- \braket{\boldsymbol a^{\dagger} \boldsymbol a +\boldsymbol b^{\dagger} \boldsymbol b + 1}^2.
    \end{split}
\end{equation}
For the family of NOON states \eqref{eq:NOONState}, we find (since the criterion contains the same moments as $d_{1,9,13}$, the calculation is completely analogous to the one presented in \hyperref[App:valueNoon]{Appendix} \ref{App:valueNoon})
\begin{equation}
    \begin{split}
        d'_{1,9,13} &= \left[ 16 \, \text{Re}^2 (\alpha^* \beta) \, \text{Im}^2 (\alpha^* \beta) - 8 \, \abs{\alpha^* \beta}^2 \right] \delta_{n 1} \\
        &\hspace{0.4cm}- 16 \, \text{Re}^2 (\alpha^* \beta) \, \delta_{n 2}.
    \end{split}
\end{equation}
Clearly, $d'_{1,9,13}$ is negative for $n=2$ and all $\alpha, \beta \in \mathbb{C}$  (which includes the aforementioned example of the Hong-Ou-Mandel state), while it is also negative for $n=1$ provided both amplitudes are for example pure real or imaginary (which includes the  aforementioned example of the first Bell state). Therefore, we may equally proceed with $d'_{1,9,13}$ instead of $d_{1,9,13}$.

We have also checked the performance of $d'_{1,9,13}$ for the mixed Schrödinger cat states \eqref{eq:CatState} considered in Sec. \ref{subsec:MixedSchrödingerCatStates}. Interestingly, the results are very similar to those of $d_{1,4,9}$ (see \hyperref[app:d1913CatStates]{Appendix} \ref{app:d1913CatStates} for a comparison). Therefore, one may choose either one of the two depending on the application: $d_{1,4,9}$ should be preferred when the measurements have to be local, while $d'_{1,9,13}$ is more useful when the number of copies should be minimized (it requires a single copy, see below).

\begin{figure}[t!]
    \centering    
    \includegraphics[width=0.45\textwidth]{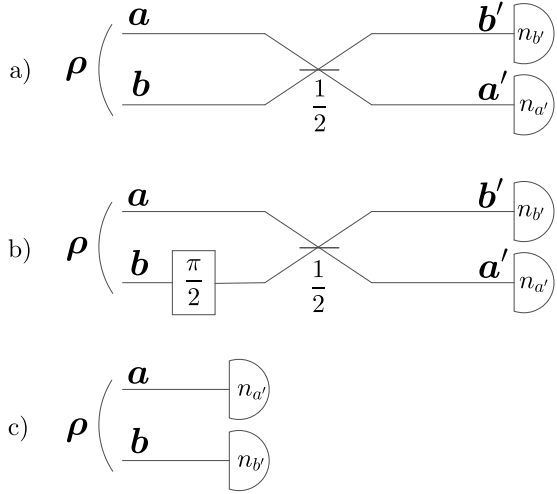}
    \caption{Optical circuits implementing the transformations $M_1$, $M_2$, and $M_3$. a) We measure $\boldsymbol{L}^x_{ab}$ by applying a balanced beam splitter between the two local modes before using photon number measurements. b) To measure $\boldsymbol{L}^y_{ab}$, we add a phase of $\frac{\pi}{2}$ on the second mode before the balanced beam splitter, followed by photon number detectors. c) $\boldsymbol{L}^z_{ab}$ directly follows from photon number measurements.}
    \label{fig:agarwal}
\end{figure}

\begin{figure*}[t!]
    \centering    
    \includegraphics[width=0.999\textwidth]{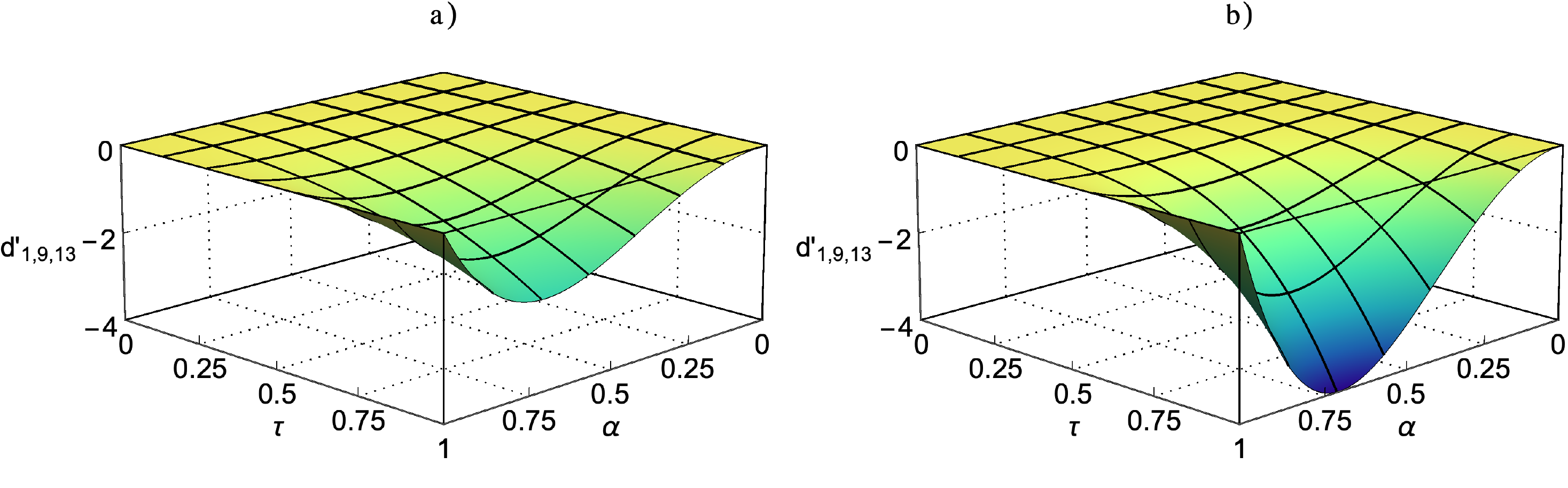}
    \caption{Entanglement witness $d'_{1,9,13}$ for NOON states as a function of transmittance $\tau$ and real amplitude $\alpha$ for $n=1$ and $2$ in a) and b), respectively. Entanglement is detected for arbitrarily small but finite losses since $d'_{1,9,13}<0$.}
    \label{fig:D1913Experiment}
\end{figure*}

\subsubsection{Multimode implementation}
Interestingly, Eq. \eqref{eq:agarwal} can be expressed in terms of spin operators across the bipartition $AB$ without the need for several copies, namely
\begin{equation}
    \label{eq:agarwalSpin}
    \begin{split}
        d'_{1,9,13} &= 16 \,\sigma^2 _{ \boldsymbol L^x_{ab}} \sigma^2_{ \boldsymbol  L^y_{ab}} + 4 \, \sigma^2_{ \boldsymbol  L_{ab}^0} - 4 \,  \sigma^2_{\boldsymbol  L_{ab}^z} \\
        & - 4\braket{\boldsymbol  L^x_{ab}}^2 - 4 \braket{\boldsymbol  L^y_{ab}}^2 - 4 \braket{\boldsymbol  L^z_{ab} }^2.
    \end{split}
\end{equation}
To access $d'_{1,9,13}$, we need three independent measurement schemes for the three spin observables $\boldsymbol{L}_{ab}^x, \boldsymbol{L}_{ab}^y, \boldsymbol{L}_{ab}^z$ (note again that $\boldsymbol{L}_{ab}^0$ can be measured simultaneously with $\boldsymbol{L}_{ab}^z$). The three corresponding transformations  
\begin{equation}
    \begin{split}
        M_1 &= \textrm{BS}_{ab}\left(1/2\right), \\
        M_2 &= \textrm{BS}_{ab}\left(1/2\right) \textrm{PS}_{b} \left(\pi/2\right), \\
        M_3 &= \textrm{I}_{ab},
    \end{split}
\end{equation}
can be respectively implemented by the three optical circuits shown in  \autoref{fig:agarwal}. These circuits transform each spin operator into $\boldsymbol{L}_{ab}^z$, whose relevant expectation values are given by
\begin{equation}
    \begin{split}
        \braket{\left(\boldsymbol{L}_{a'b'}^z \right)^2} &= \frac{1}{4} \left( \braket{\boldsymbol{n}_{a'}^2} - 2 \braket{\boldsymbol{n}_{a'} \boldsymbol{n}_{b'}} + \braket{\boldsymbol{n}_{b'}^2} \right), \\
        \braket{\boldsymbol{L}_{a'b'}^z}^2 &= \frac{1}{4} \left( \braket{\boldsymbol{n}_{a'}}^2 - 2 \braket{\boldsymbol{n}_{a'}} \braket{\boldsymbol{n}_{b'}} + \braket{\boldsymbol{n}_{b'}}^2 \right).
    \end{split}
\end{equation}
Thus, applying these circuits and measuring the photon numbers yields the needed mean values and variances, so we obtain $d'_{1,9,13}$ using Eq. \eqref{eq:agarwalSpin}.

Let us remark that, compared to the other two entanglement witnesses  discussed in \autoref{subsec:GaussianStates} and \autoref{subsec:MixedSchrödingerCatStates} where Alice and Bob had to count photons locally on their copies, $d'_{1,9,13} < 0$ is a non-local condition in the sense that Alice and Bob have to perform interferometric measurements on their joint system $AB$. 

\subsubsection{Imperfect copies and optical losses}

We do not need to analyze the effect of imperfect copies on $d_{1,9,13}$ since we have not developed a multicopy implementation of it. Nevertheless, it is worth illustrating the fact that this criterion does not suffer from false positives by considering the value of $d_{1,9,13}$ when inputting three different product states consisting each of two Fock states. For such states, all off-diagonal elements of the matrix vanish, so that the determinant is simply the product of the diagonal elements, which are all positive. The determinant is thus always positive and there are no false-positive detections.

Now coming to the criterion based on $d'_{1,9,13}$, analyzing imperfect copies is meaningless since there is no need to use more than one copy to measure it. We can only analyze the effect of losses. Adding losses to the two inputs of the optical circuit leads to the expression
\begin{equation}
    \begin{split}
        d'_{1,9,13} &= \Big( 16 \, \text{Re}^2 (\alpha^* \beta)\,  \text{Im}^2 (\alpha^* \beta)\, \tau_a^2 \tau_b^2 \\
        &-4 \, (\abs{\alpha}^2 \tau_a +\abs{\beta}^2 \tau_b+1)\, \tau_a \tau_b \abs{\alpha^* \beta}^2 \Big) \delta_{n 1} \\
        &- 16 \, \text{Re}^2 (\alpha^* \beta) \tau_a^2 \tau_b^2 \delta_{n 2}.
    \end{split}
\end{equation}
We exemplify the dependence on the transmittance $\tau \equiv \tau_a = \tau_b$ for the special case where $\alpha$ and $\beta$ are real in \autoref{fig:D1913Experiment}a and \autoref{fig:D1913Experiment}b for $n=1$ and $2$, respectively. In the former case, the (negative) value of $d'_{1,9,13}$ increases cubically with $\tau$, while in the latter case it increases quartically with $\tau$. In both cases, entanglement is detected for all amplitudes $\alpha \neq 0,1$, and non-zero transmittance $\tau > 0$, with the violation of the separability criterion being the largest around $\alpha \approx 3/4$.

\section{Conclusion and Outlook}
\label{sec:Conclusion}
To summarize, we have put forward schemes to efficiently access three continuous-variable separability criteria based on multimode operators, which are read out via linear interferometers and photon number measurements. The implementation of these schemes thus requires interferometric stability over the few replicas of the state of interest as well as photon-number resolving detectors. The benefit is that the separability criteria are directly accessed, implying that state tomography is not needed. Our schemes encompass optical circuits for second-moment criteria to detect entanglement of Gaussian states, as well as two types of fourth-order criteria suitable for witnessing entanglement in case of mixed Schrödinger cat states (for full parameter ranges) and NOON states (for low-energetic excitations), respectively.

While we focused on three specific separability criteria, our approach is in no way limited to those. Hence, it is of particular interest to identify other sets of relevant criteria and devise suitable multimode observables and corresponding measurement schemes. For example, one may investigate other prominent second-order criteria such as the Simon criterion \cite{Simon2000}, which is equivalent to the condition $d_{1,2,3,4,5} \ge 0$ \cite{Shchukin2005}, such that a multicopy implementation would require five replicas (or four if the invariance under displacements is exploited). Alternatively, one may try to implement the second-order criteria due to Mancini \textit{et al.} \cite{Mancini2002,Giovannetti2003}, which constrain the product of the variances appearing in \eqref{eq:DuanCriteria} instead of their sum. Both criteria are interesting as they are stronger than the criteria by Duan \textit{et al.} \cite{Duan2000} as well as the condition $d_{1,2,4} \ge 0$ (all are equivalent in the Gaussian case).

Furthermore, given that our method is generic and based on the algebraic properties of spin operators, a more systematic approach, especially for more than three copies, would be eligible. This may lead to feasible multicopy observables beyond three copies, which could allow us to formulate multicopy versions of entanglement witnesses beyond fourth-order moments. In addition, the method should be equally applicable to other bosonic systems characterized by the pair $\boldsymbol{a},\boldsymbol{a}^{\dagger}$ satisfying $[\boldsymbol{a},\boldsymbol{a}^{\dagger}] = 1$, going beyond quantum optics.

At last, let us remark that the experimental application of our schemes is within reach of current technologies. As a matter of fact, the multicopy nonclassicality observable presented in Ref. \cite{Griffet2022b} has been successfully accessed on a cloud quantum computer in a recent experiment \cite{Goldberg2023}, thereby suggesting the general feasibility of the multicopy method. As we have shown here, the typical experimental imperfections should have a modest influence on the detection of entanglement. All multimode observables that we have analyzed are robust against losses in the sense that finite losses decrease the chances for entanglement detection but never completely prevent it. In all cases, false-positive detection of entanglement could be excluded. Nevertheless, a deeper analysis of experimental imperfections, for instance noise effects and finite detector resolution, would be valuable towards an experimental implementation of our method.

\section*{Acknowledgements}
C.G. is a Research Fellow of the Fonds de la Recherche Scientifique - FNRS. N.J.C. acknowledges support from the Fonds de la Recherche Scientifique - FNRS under Grant No. T.0224.18. T.H. and N.J.C. are supported by the European Union under project ShoQC within the ERA-NET Cofund in Quantum Technologies (QuantERA) program.
\begin{appendix}

\section{Comparing $d_{1,2,4}<0$ with the criterion of Duan \textit{et al.}}
\label{app:DuanCriterion}
One can show that $d_{1,2,4}$ reduces to
\begin{equation}
    d_{1,2,4} = \sigma_{\boldsymbol{a}^{\dagger} \boldsymbol{a}} \sigma_{\boldsymbol{b}^{\dagger} \boldsymbol{b}} - \sigma_{\boldsymbol{a}^{\dagger} \boldsymbol{b}^{\dagger}} \sigma_{\boldsymbol{ab}},
    \label{eq:d124ModeOperators}
\end{equation}
where
\begin{equation}
    \sigma_{\boldsymbol{y z}} = \braket{\boldsymbol{y z}} - \braket{\boldsymbol{y}} \braket{\boldsymbol{z}}
    \label{eq:CovarianceDefinition}
\end{equation}
denotes the covariance of the two observables $\boldsymbol{y},\boldsymbol{z}$. By employing the identities
\begin{equation}
    \begin{split}
        \sigma_{\boldsymbol{a}^{\dagger} \boldsymbol{a}} &= \frac{1}{2} \left( \sigma_{\boldsymbol{x}_1}^2 + \sigma_{\boldsymbol{p}_1}^2 - 1  \right), \\
        \sigma_{\boldsymbol{b}^{\dagger} \boldsymbol{b}} &= \frac{1}{2} \left( \sigma_{\boldsymbol{x}_2}^2 + \sigma_{\boldsymbol{p}_2}^2 - 1  \right), \\
        \sigma_{\boldsymbol{a} \boldsymbol{b}} &= \frac{1}{2} \left( \sigma_{\boldsymbol{x}_1 \boldsymbol{x}_1} + i \sigma_{\boldsymbol{x}_1 \boldsymbol{p}_2} + i \sigma_{\boldsymbol{x}_2 \boldsymbol{p}_1} - \sigma_{\boldsymbol{p}_1 \boldsymbol{p}_1} \right),\\
        \sigma_{\boldsymbol{a}^{\dagger} \boldsymbol{b}^{\dagger}} &= \sigma_{\boldsymbol{a} \boldsymbol{b}}^{\dagger},
    \end{split}
\end{equation}
the condition $d_{1,2,4} \ge 0$ can be translated into a condition on the local quadratures and their correlations
\begin{equation}
    \begin{split}
        0 \le &\left( \sigma_{\boldsymbol{x}_1}^2 + \sigma_{\boldsymbol{p}_1}^2 - 1  \right) \left( \sigma_{\boldsymbol{x}_2}^2 + \sigma_{\boldsymbol{p}_2}^2 - 1  \right) \\
        &-\left(\sigma_{\boldsymbol{x}_1 \boldsymbol{x}_1} - \sigma_{\boldsymbol{p}_1 \boldsymbol{p}_1} \right)^2 -\left(\sigma_{\boldsymbol{x}_1 \boldsymbol{p}_2} + \sigma_{\boldsymbol{x}_2 \boldsymbol{p}_1} \right)^2.
    \end{split}
    \label{eq:d124LocalQuadratures}
\end{equation}
Similarly, rewriting the criterion \eqref{eq:DuanCriteria} in terms of local quadratures using
\begin{equation}
    \begin{split}
        \sigma^2_{\boldsymbol{x}_{\pm}} &= r^2 \sigma^2_{\boldsymbol{x}_1} + \frac{1}{r^2} \sigma^2_{\boldsymbol{x}_2} \pm 2 \sigma_{\boldsymbol{x}_1 \boldsymbol{x}_2}, \\
        \sigma^2_{\boldsymbol{p}_{\pm}} &= r^2 \sigma^2_{\boldsymbol{p}_1} + \frac{1}{r^2} \sigma^2_{\boldsymbol{p}_2} \pm 2 \sigma_{\boldsymbol{p}_1 {p}_2},
    \end{split}
\end{equation}
allows one to optimize over $r$ by searching for a global minimum. One finds
\begin{equation}
    r^2 = \sqrt{\frac{\sigma_{\boldsymbol{x}_2}^2 + \sigma_{\boldsymbol{p}_2}^2 - 1}{\sigma_{\boldsymbol{x}_1}^2 + \sigma_{\boldsymbol{p}_1}^2 - 1}},
\end{equation}
such that the optimal Duan criterion in local variables reads
\begin{equation}
    \begin{split}
        d_{\text{Duan}} &= 2 \sqrt{\left( \sigma_{\boldsymbol{x}_1}^2 + \sigma_{\boldsymbol{p}_1}^2 - 1 \right) \left( \sigma_{\boldsymbol{x}_2}^2 + \sigma_{\boldsymbol{p}_2}^2 - 1 \right)} \\
        &\hspace{0.4cm}\pm 2 \left(\sigma_{\boldsymbol{x}_1 \boldsymbol{x}_2} - \sigma_{\boldsymbol{p}_1 \boldsymbol{p}_2} \right).
    \end{split}
    \label{eq:DuanCriteriaOptimal}
\end{equation}
The non-negativity of the latter is equivalent to the condition
\begin{equation}
    \begin{split}
        0 \le &\left( \sigma_{\boldsymbol{x}_1}^2 + \sigma_{\boldsymbol{p}_1}^2 - 1  \right) \left( \sigma_{\boldsymbol{x}_2}^2 + \sigma_{\boldsymbol{p}_2}^2 - 1  \right) \\
        &-\left(\sigma_{\boldsymbol{x}_1 \boldsymbol{x}_1} - \sigma_{\boldsymbol{p}_1 \boldsymbol{p}_1} \right)^2.
    \end{split}
    \label{eq:DuanCriteriaOptimal2}
\end{equation}
By comparing \eqref{eq:d124LocalQuadratures} and \eqref{eq:DuanCriteriaOptimal2} it becomes apparent that $d_{1,2,4} \ge 0$ implies $d_{\text{Duan}} \ge 0$ since $\left(\sigma_{\boldsymbol{x}_1 \boldsymbol{p}_2} + \sigma_{\boldsymbol{x}_2 \boldsymbol{p}_1} \right)^2 \ge 0$. Therefore, $d_{1,2,4} \ge 0$ is stronger than $d_{\text{Duan}} \ge 0$ in the sense that the former condition contains additional information about the correlations between quadratures of different types.

\section{Calculation of the determinants for several classes of states}
\subsection{$d_{1,2,4}$ for the two-mode squeezed vacuum state}
\label{App:valueTMSV}
To evaluate the determinant $d_{1,2,4}$ (Eq. \eqref{eq:d124}) for the two-mode squeezed vacuum state \eqref{eq:TMSVState}, we use the lowering or raising property of the annihilation or creation operator
\begin{equation}
    \begin{split}
        \boldsymbol a \ket{n,n} &= \sqrt{n} \ket{n-1,n}, \\ \boldsymbol a^\dagger \ket{n,n} &= \sqrt{n+1} \ket{n+1,n},
        \label{eq:RaisingLowering}
    \end{split}
\end{equation}
and similarly for the creation and annihilation operators acting on mode $\boldsymbol b$. Then follows for the first non-trivial matrix element of $d_{1, 2,4}$
\begin{equation}
    \begin{split} 
        \braket{\boldsymbol a^\dagger \boldsymbol a} &= (1-\lambda^2)\sum_{n,n'=0}^\infty \lambda^{n} \lambda^{n'} \bra{n',n'} \boldsymbol a^\dagger \boldsymbol a \ket{n,n} \\
        &= (1-\lambda^2)\sum_{n,n'=0}^\infty \lambda^{n}\lambda^{n'} \sqrt{n n'} \, \delta_{n n'} \\
        &= (1-\lambda^2)\sum_{n=0}^\infty \lambda^{2 n} n \\
        &= \frac{\lambda^2}{1-\lambda^2},
    \end{split}
\end{equation}
where we used the orthonormality of Fock states $\braket{n | n'} = \delta_{n n'}$. The remaining matrix elements are found analogously, leading to the expression \eqref{eq:d124TMSV} for the determinant
\begin{equation}
    d_{1,2,4} = \begin{vmatrix}
    1 & 0 & 0\\
    0 & \frac{\lambda^2}{1-\lambda^2} & \frac{\lambda}{1-\lambda^2} \\
    0 & \frac{\lambda}{1-\lambda^2}  & \frac{\lambda^2}{1-\lambda^2} 
    \end{vmatrix} = - \frac{\lambda^2}{1 - \lambda^2}.
\end{equation}

\subsection{$d_{1,4,9}$ for mixed Schrödinger cat states}\label{App:valueCat}
We calculate the value of the determinant $d_{1,4,9}$ [Eq. \eqref{eq:d149}] for general entangled Schrödinger cat states defined in Eq. \eqref{eq:CatState} by using that canonical coherent states are eigenstates of the annihilation operator
\begin{equation}
    \begin{split}
        \boldsymbol{a} \ket{\alpha, \beta} &= \alpha \ket{\alpha, \beta}, \\
        \bra{\alpha, \beta} \boldsymbol{a}^{\dagger} &= \bra{\alpha, \beta} \alpha^*, 
    \end{split}
\end{equation}
and similarly for mode $B$. We start with the matrix element $\braket{\boldsymbol b^\dagger}$, which evaluates to
\begin{equation}
    \begin{split}
        \braket{\boldsymbol b^\dagger} &\propto \Tr\Big[\ket{\alpha, \beta} \bra{\alpha, \beta} \boldsymbol b^\dagger + \ket{-\alpha, -\beta} \bra{-\alpha, -\beta} \boldsymbol b^\dagger\\
        &- (1-z)  \left( \ket{\alpha, \beta} \bra{-\alpha, -\beta} \boldsymbol b^\dagger+ \ket{-\alpha, -\beta} \bra{\alpha, \beta} \boldsymbol b^\dagger \right) \Big],\\
        &= \Tr \Big[\ket{\alpha, \beta} \bra{\alpha, \beta}  \beta^* + \ket{-\alpha, -\beta} \bra{-\alpha, -\beta} (-\beta^*)\\
        &+ (1-z)  \left( \ket{\alpha, \beta} \bra{-\alpha, -\beta}\beta^* - \ket{-\alpha, -\beta} \bra{\alpha, \beta}  \beta^* \right) \Big],\\
        &= 0.
    \end{split}
\end{equation}
Analogously, we find for the remaining matrix elements
\begin{equation}
    \begin{split}
    &\braket{\boldsymbol{b}} = \braket{\boldsymbol{b}^{\dagger}} = 0,\\
    &\braket{\boldsymbol{a} \boldsymbol{b}^{\dagger}}= 2 \alpha \beta^* N(\alpha,\beta,z) (1+(1-z)e^{-2\abs{\alpha}^2-2\abs{\beta}^2}),\\
    &\braket{\boldsymbol{a}^{\dagger} \boldsymbol{b}} = 2 \alpha^* \beta N(\alpha,\beta,z) (1+(1-z)e^{-2\abs{\alpha}^2-2\abs{\beta}^2}),\\
    &\braket{\boldsymbol{b}^{\dagger} \boldsymbol{b}} = 2 \abs{\beta}^2 N(\alpha,\beta,z) (1+(1-z)e^{-2\abs{\alpha}^2-2\abs{\beta}^2}),\\
    &\braket{\boldsymbol{a} \boldsymbol{b}^{\dagger} \boldsymbol{b}} = \braket{\boldsymbol{a}^{\dagger} \boldsymbol{b}^{\dagger} \boldsymbol{b}} = 0,\\
    &\braket{\boldsymbol{a}^{\dagger} \boldsymbol{a} \boldsymbol{b}^{\dagger} \boldsymbol{b}} = \abs{\alpha}^2 \abs{\beta}^2.
    \end{split}
\end{equation}
The full determinant of $d_{1,4,9}$ given in Eq. \eqref{eq:d149cat} follows then after identifying the hyperbolic functions
\begin{equation}
    \begin{split}
        &\coth \left[ \abs{\alpha}^2 + \abs{\beta}^2 - \frac{1}{2} \ln \left(1-z \right) \right] \\
        &= 2 N (\alpha, \beta, z) \left[ 1 + (1-z) e^{-2 \left( \abs{\alpha}^2 + \abs{\beta}^2 \right)} \right],
    \end{split}
\end{equation}
and
\begin{equation}
    \begin{split}
        &\sinh^{-2} \left[ \abs{\alpha}^2 + \abs{\beta}^2 - \frac{1}{2} \ln \left(1-z \right) \right] \\
        &= 4 N^2 (\alpha, \beta, z) \left[ 1 + (1-z) e^{-2 \left( \abs{\alpha}^2 + \abs{\beta}^2 \right)} \right]^2 - 1.
    \end{split}
\end{equation}

\subsection{$d_{1,9,13}$ for NOON states}\label{App:valueNoon}

In order to evaluate $d_{1,9,13}$ [Eq. \eqref{eq:d1913}], we will use again the properties of the creation and annihilation operators given in \eqref{eq:RaisingLowering}. For example, for the matrix element $\braket{\boldsymbol a \boldsymbol{b}^{\dagger}}$ we find
\begin{equation}
    \begin{split}
        \braket{\boldsymbol a \boldsymbol{b}^{\dagger}} &= \Big(  \alpha^* \bra{n,0} + \beta^* \bra{0,n} \Big)\ \boldsymbol a \boldsymbol{b}^{\dagger} \Big(  \alpha \ket{n,0} + \beta \ket{0,n} \Big),\\
        &= \Big(  \alpha^* \bra{n,0} + \beta^* \bra{0,n} \Big) \sqrt{n} \alpha \ket{n-1,1}, \\
        &=  \beta^*   \sqrt{n} \alpha\delta_{n1}, \\
        &= \alpha \beta^* \delta_{n1}.
    \end{split}
\end{equation}
Repeating this strategy for the other matrix elements leads to the determinant
\begin{equation}
    \begin{split}
        d_{1,9,13} &= \begin{vmatrix}
        1 & \alpha \beta^* \delta_{n1} & \alpha^* \beta \delta_{n1}\\
        \alpha^* \beta \delta_{n1} & 0 & 2 \alpha^* \beta \delta_{n2} \\
        \alpha \beta^* \delta_{n1} & 2 \alpha \beta^* \delta_{n2} & n+1
        \end{vmatrix} \\
        &=- 2 \abs{\alpha}^2 \abs{\beta}^2 \left( \delta_{n 1} + 2 \delta_{n 2} \right),
    \end{split}
\end{equation}
which is nothing but Eq. \eqref{eq:d1913Noon}.

\newpage

\section{Invariance of $d_{1,2,4}$ under displacements}
\label{app:Displacementd124}
We prove that applying an arbitrary displacement $\boldsymbol D (\alpha) \boldsymbol D (\beta)$ on the bipartite state $\boldsymbol \rho$ does not change the value of the determinant $d_{1,2,4}$. The annihilation operators $\boldsymbol a$ and $\boldsymbol b$ of the two subsystems transform as
\begin{equation}
    \begin{split}
        \boldsymbol a & \rightarrow \boldsymbol{a}' = \boldsymbol a + \alpha,\\
        \boldsymbol b & \rightarrow \boldsymbol{b}' = \boldsymbol b + \beta,
    \end{split}
    \label{eq:DisplacementTransfo}
\end{equation}
with complex phases $\alpha, \beta \in \mathbb{C}$. Then, the determinant transforms as
\begin{widetext}
    \begin{equation}
        \begin{split}
            d_{1,2,4} \to d'_{1,2,4} &=
            \begin{vmatrix}
            1 & \braket{\boldsymbol{a}'} & \braket{\boldsymbol{b}'^{\dagger}} \\
            \braket{\boldsymbol{a}'^{\dagger}} & \braket{\boldsymbol{a}'^{\dagger}\boldsymbol{a}'} & \braket{\boldsymbol{a}'^{\dagger}\boldsymbol{b}'^\dagger} \\
            \braket{\boldsymbol{b}'} & \braket{\boldsymbol{a}' \boldsymbol{b}'} & \braket{\boldsymbol{b}'^{\dagger} \boldsymbol{b}'}
            \end{vmatrix} \\
            &= \begin{vmatrix}
            1 & \braket{\boldsymbol{a}+\alpha} & \braket{\boldsymbol{b}^{\dagger}+\beta^*}\\
            \braket{\boldsymbol{a}^{\dagger}+ \alpha^*} & \braket{(\boldsymbol{a}^{\dagger}+\alpha^*)(\boldsymbol a+\alpha)} & \braket{(\boldsymbol{a}^{\dagger}+\alpha^*)(\boldsymbol{b}^{\dagger}+\beta^*)}\\
            \braket{\boldsymbol b +\beta} & \braket{(\boldsymbol{a }+\alpha)( \boldsymbol{b}+\beta)} & \braket{(\boldsymbol{b}^{\dagger}+\beta^*)( \boldsymbol{b}+\beta)}
            \end{vmatrix} \\
            &= \begin{vmatrix}
            1 & \braket{\boldsymbol{a}} + \alpha& \braket{\boldsymbol{b}^{\dagger}} + \beta^*\\
            \braket{\boldsymbol{a}^{\dagger}} + \alpha^* & \braket{\boldsymbol{a}^{\dagger}\boldsymbol a} + \alpha \braket{\boldsymbol{a}^{\dagger}} + \alpha^* (\braket{\boldsymbol{a}} + \alpha) & \braket{\boldsymbol{a}^{\dagger}\boldsymbol b^\dagger} + \beta^* \braket{\boldsymbol{a}^{\dagger}} + \alpha^* ( \braket{\boldsymbol{b}^{\dagger}} + \beta^*)\\
            \braket{\boldsymbol b} + \beta & \braket{\boldsymbol{a b}} + \alpha \braket{\boldsymbol b} + \beta (\braket{\boldsymbol{a}} + \alpha)& \braket{\boldsymbol{b}^{\dagger} \boldsymbol{b}} + \beta^*\braket{\boldsymbol b} + \beta (\braket{\boldsymbol{b}^{\dagger}} + \beta^*)
            \end{vmatrix} \\
            &= \begin{vmatrix}
            1 & \braket{\boldsymbol{a}} + \alpha& \braket{\boldsymbol{b}^{\dagger}} + \beta^*\\
            \braket{\boldsymbol{a}^{\dagger}} & \braket{\boldsymbol{a}^{\dagger}\boldsymbol a} + \alpha \braket{\boldsymbol{a}^{\dagger}} & \braket{\boldsymbol{a}^{\dagger}\boldsymbol b^\dagger} + \beta^* \braket{\boldsymbol{a}^{\dagger}}\\
            \braket{\boldsymbol b} & \braket{\boldsymbol{a b}} + \alpha \braket{\boldsymbol b}& \braket{\boldsymbol{b}^{\dagger} \boldsymbol{b}} + \beta^*\braket{\boldsymbol b}
            \end{vmatrix} \\
            &= d_{1,2,4},
        \end{split}
    \end{equation}
where we used that the determinant remains invariant when adding to a column or row another column or row multiplied by some complex number in the two last equations.

\clearpage

\section{Decomposition of $d_{2,4}$}
\label{app:B24Details}
We start from the four products $\boldsymbol{L}^j_{a_{2'3'}} \boldsymbol{L}^j_{b_{2'3'}}$ for $j \in \{x,y,z,0\}$ applied to the modes $2'$ and $3'$, which read
\begin{equation}
    \begin{split}
        \boldsymbol{L}_{a_{2'3'}}^x \boldsymbol{L}_{b_{2'3'}}^x &= \frac{1}{4} \Big(\boldsymbol a^{\dagger}_{3'} \boldsymbol a_{2'} \boldsymbol b^{\dagger}_{3'} \boldsymbol b_{2'} + \boldsymbol a^{\dagger}_{2'} \boldsymbol a_{3'} \boldsymbol b^{\dagger}_{2'} \boldsymbol b_{3'} +\boldsymbol a^{\dagger}_{2'} \boldsymbol a_{3'} \boldsymbol b^{\dagger}_{3'} \boldsymbol b_{2'}+\boldsymbol a^{\dagger}_{3'} \boldsymbol a_{2'} \boldsymbol b^{\dagger}_{2'} \boldsymbol b_{3'} \Big), \\
        \boldsymbol{L}_{a_{2'3'}}^y \boldsymbol{L}_{b_{2'3'}}^y &= - \frac{1}{4} \Big(\boldsymbol a^{\dagger}_{3'} \boldsymbol a_{2'} \boldsymbol b^{\dagger}_{3'} \boldsymbol b_{2'}+\boldsymbol a^{\dagger}_{2'} \boldsymbol a_{3'} \boldsymbol b^{\dagger}_{2'} \boldsymbol b_{3'} -\boldsymbol a^{\dagger}_{2'} \boldsymbol a_{3'} \boldsymbol b^{\dagger}_{3'} \boldsymbol b_{2'}-\boldsymbol a^{\dagger}_{3'} \boldsymbol a_{2'} \boldsymbol b^{\dagger}_{2'} \boldsymbol b_{3'} \Big), \\
        \boldsymbol{L}_{a_{2'3'}}^z \boldsymbol{L}_{b_{2'3'}}^z &= \frac{1}{4} \Big(\boldsymbol a^{\dagger}_{2'} \boldsymbol a_{2'} \boldsymbol b^{\dagger}_{2'} \boldsymbol b_{2'}+\boldsymbol a^{\dagger}_{3'} \boldsymbol a_{3'} \boldsymbol b^{\dagger}_{3'} \boldsymbol b_{3'} -\boldsymbol a^{\dagger}_{2'} \boldsymbol a_{2'} \boldsymbol b^{\dagger}_{3'} \boldsymbol b_{3'}-\boldsymbol a^{\dagger}_{3'} \boldsymbol a_{3'} \boldsymbol b^{\dagger}_{2'} \boldsymbol b_{2'} \Big), \\
        \boldsymbol{L}_{a_{2'3'}}^0 \boldsymbol{L}_{b_{2'3'}}^0 &= \frac{1}{4} \Big(\boldsymbol a^{\dagger}_{2'} \boldsymbol a_{2'} \boldsymbol b^{\dagger}_{2'} \boldsymbol b_{2'}+\boldsymbol a^{\dagger}_{3'} \boldsymbol a_{3'} \boldsymbol b^{\dagger}_{3'} \boldsymbol b_{3'} +\boldsymbol a^{\dagger}_{2'} \boldsymbol a_{2'} \boldsymbol b^{\dagger}_{3'} \boldsymbol b_{3'}+\boldsymbol a^{\dagger}_{3'} \boldsymbol a_{3'} \boldsymbol b^{\dagger}_{2'} \boldsymbol b_{2'} \Big).\\
    \end{split}
\end{equation}
Then, following the definitions of the operators $\boldsymbol C_j$ given in \eqref{eq:DefinitionA} we find
\begin{equation}
    \begin{split}
        \boldsymbol C_1 &=\frac{1}{4} \Big(\boldsymbol a^{\dagger}_{2'} \boldsymbol a_{2'} \boldsymbol b^{\dagger}_{2'} \boldsymbol b_{2'} + \boldsymbol a^{\dagger}_{3'} \boldsymbol a_{3'} \boldsymbol b^{\dagger}_{3'} \boldsymbol b_{3'} + \boldsymbol a^{\dagger}_{2'} \boldsymbol a_{2'} \boldsymbol b^{\dagger}_{3'} \boldsymbol b_{3'} +\boldsymbol a^{\dagger}_{3'} \boldsymbol a_{3'} \boldsymbol b^{\dagger}_{2'} \boldsymbol b_{2'} \\
        &\hspace{0.6cm}- \boldsymbol a^{\dagger}_{3'} \boldsymbol a_{2'} \boldsymbol b^{\dagger}_{3'} \boldsymbol b_{2'} - \boldsymbol a^{\dagger}_{2'} \boldsymbol a_{3'} \boldsymbol b^{\dagger}_{2'} \boldsymbol b_{3'} -\boldsymbol a^{\dagger}_{2'} \boldsymbol a_{3'} \boldsymbol b^{\dagger}_{3'} \boldsymbol b_{2'}-\boldsymbol a^{\dagger}_{3'} \boldsymbol a_{2'} \boldsymbol b^{\dagger}_{2'} \boldsymbol b_{3'} \Big), \\
        \boldsymbol C_2 &=\frac{1}{4} \Big(\boldsymbol a^{\dagger}_{2'} \boldsymbol a_{2'} \boldsymbol b^{\dagger}_{2'} \boldsymbol b_{2'} + \boldsymbol a^{\dagger}_{3'} \boldsymbol a_{3'} \boldsymbol b^{\dagger}_{3'} \boldsymbol b_{3'} + \boldsymbol a^{\dagger}_{2'} \boldsymbol a_{2'} \boldsymbol b^{\dagger}_{3'} \boldsymbol b_{3'} + \boldsymbol a^{\dagger}_{3'} \boldsymbol a_{3'} \boldsymbol b^{\dagger}_{2'} \boldsymbol b_{2'} \\
        &\hspace{0.6cm}+ \boldsymbol a^{\dagger}_{3'} \boldsymbol a_{2'} \boldsymbol b^{\dagger}_{3'} \boldsymbol b_{2'} + \boldsymbol a^{\dagger}_{2'} \boldsymbol a_{3'} \boldsymbol b^{\dagger}_{2'} \boldsymbol b_{3'} - \boldsymbol a^{\dagger}_{2'} \boldsymbol a_{3'} \boldsymbol b^{\dagger}_{3'} \boldsymbol b_{2'} - \boldsymbol a^{\dagger}_{3'} \boldsymbol a_{2'} \boldsymbol b^{\dagger}_{2'} \boldsymbol b_{3'} \Big), \\
        \boldsymbol C_3 &=\frac{1}{2} \Big(\boldsymbol a^{\dagger}_{2'} \boldsymbol a_{2'} \boldsymbol b^{\dagger}_{3'} \boldsymbol b_{3'}+\boldsymbol a^{\dagger}_{3'} \boldsymbol a_{3'} \boldsymbol b^{\dagger}_{2'} \boldsymbol b_{2'} \Big),
    \end{split}
\end{equation}
such that
\begin{equation}
    \begin{split}
        \boldsymbol{C}_1 - \boldsymbol{C}_2 + \boldsymbol{C}_3 &= \frac{1}{2} \Big(\boldsymbol{a}^{\dagger}_{2'} \boldsymbol{a}_{2'} \boldsymbol{b}^{\dagger}_{3'} \boldsymbol{b}_{3'} + \boldsymbol{a}^{\dagger}_{3'} \boldsymbol{a}_{3'} \boldsymbol{b}^{\dagger}_{2'} \boldsymbol{b}_{2'} -\boldsymbol{a}^{\dagger}_{2'} \boldsymbol{a}_{3'} \boldsymbol{b}^{\dagger}_{2'} \boldsymbol{b}_{3'} -\boldsymbol{a}^{\dagger}_{3'} \boldsymbol{a}_{2'} \boldsymbol{b}^{\dagger}_{3'} \boldsymbol{b}_{2'} \Big) \\
        &= \boldsymbol D_{2,4}.
    \end{split}
\end{equation}

\section{Imperfect copies and optical losses for $d_{1,4,9}$}
\label{app:d149imperfections}
For three distinct copies of a mixed Schrödinger cat state \eqref{eq:CatState} at the input and with losses incorporated, we find for the corresponding multicopy expectation value
\begin{equation}
    \begin{split}
        \braket{\braket{\braket{\boldsymbol{D}_{1,4,9}}}} &=\frac{1}{3}\sum_{\sigma \in P_{123}} \tau_{b_{\sigma(1)}}\abs{\beta_{\sigma(1)}}^2 N(\alpha_{\sigma(1)},\beta_{\sigma(1)},z_{\sigma(1)})
        \left(1+(1-z_{\sigma(1)}) e^{-2 \abs{\alpha_{\sigma(1)}}^2-2 \abs{\beta_{\sigma(1)}}^2})\right)\\
        &\Big [ \tau_{a_{\sigma(2)}}\tau_{b_{\sigma(2)}} \abs{\alpha_{\sigma(2)}}^2 \abs{\beta_{\sigma(2)}}^2+\tau_{a_{\sigma(3)}}\tau_{b_{\sigma(3)}} \abs{\alpha_{\sigma(3)}}^2 \abs{\beta_{\sigma(3)}}^2
        -4\sqrt{\tau_{a_{\sigma(2)}}\tau_{b_{\sigma(2)}}\tau_{a_{\sigma(3)}}\tau_{b_{\sigma(3)}}}
        \\&N(\alpha_{\sigma(2)},\beta_{\sigma(2)},z_{\sigma(2)}) N(\alpha_{\sigma(3)},\beta_{\sigma(3)},z_{\sigma(3)})
        \left(1+(1-z_{\sigma(2)}) e^{-2 \abs{\alpha_{\sigma(2)}}^2-2 \abs{\beta_{\sigma(2)}}^2}\right)
        \\&\left(1+(1-z_{\sigma(3)}) e^{-2 \abs{\alpha_{\sigma(3)}}^2-2 \abs{\beta_{\sigma(3)}}^2}\right)
        \left(\alpha_{\sigma(2)}\beta_{\sigma(2)}^*\alpha_{\sigma(3)}^*\beta_{\sigma(3)}+\alpha_{\sigma(2)}^*\beta_{\sigma(2)}\alpha_{\sigma(3)}\beta_{\sigma(3)}^*\right)\Big].
    \end{split}
\end{equation}

\section{Multicopy observable associated to $d_{1,9,13}$}
\label{app:d1913}
The standard route to access $d_{1,9,13}$ is to define the multicopy observable
\begin{equation}
    \begin{split}
        \boldsymbol D_{1,9,13} = \frac{1}{\abs{S_{123}}} \sum_{\sigma \in S_{123}}
        \begin{vmatrix}
        1 & \boldsymbol a_{\sigma(1)} \boldsymbol{b}^{\dagger}_{\sigma(1)} & \boldsymbol{a}^{\dagger}_{\sigma(1)} \boldsymbol{b}_{\sigma(1)}\\
        \boldsymbol a^\dagger_{\sigma(2)} \boldsymbol{b}_{\sigma(2)} & \boldsymbol{a}^{\dagger}_{\sigma(2)} \boldsymbol{a}_{\sigma(2)} \boldsymbol{b}^{\dagger}_{\sigma(2)} \boldsymbol{b}_{\sigma(2)} & \boldsymbol{a}^{\dagger 2}_{\sigma(2)} \boldsymbol{b}^{2}_{\sigma(2)} \\
        \boldsymbol{a}_{\sigma(3)} \boldsymbol{b}^{\dagger}_{\sigma(3)} & \boldsymbol{a}^{2}_{\sigma(3)} \boldsymbol{b}^{\dagger 2}_{\sigma(3)}  & \boldsymbol{a}_{\sigma(3)} \boldsymbol{a}^{\dagger}_{\sigma(3)} \boldsymbol{b}_{\sigma(3)} \boldsymbol{b}^{\dagger}
        \end{vmatrix},
    \end{split}
\end{equation}
with $d_{1,9,13} = \braket{\braket{\braket{\boldsymbol{D}_{1,9,13}}}}$. Writing this operator in terms of spin operators leads to
\begin{figure*}[t!]
	\centering    
	\includegraphics[width=0.999\textwidth]{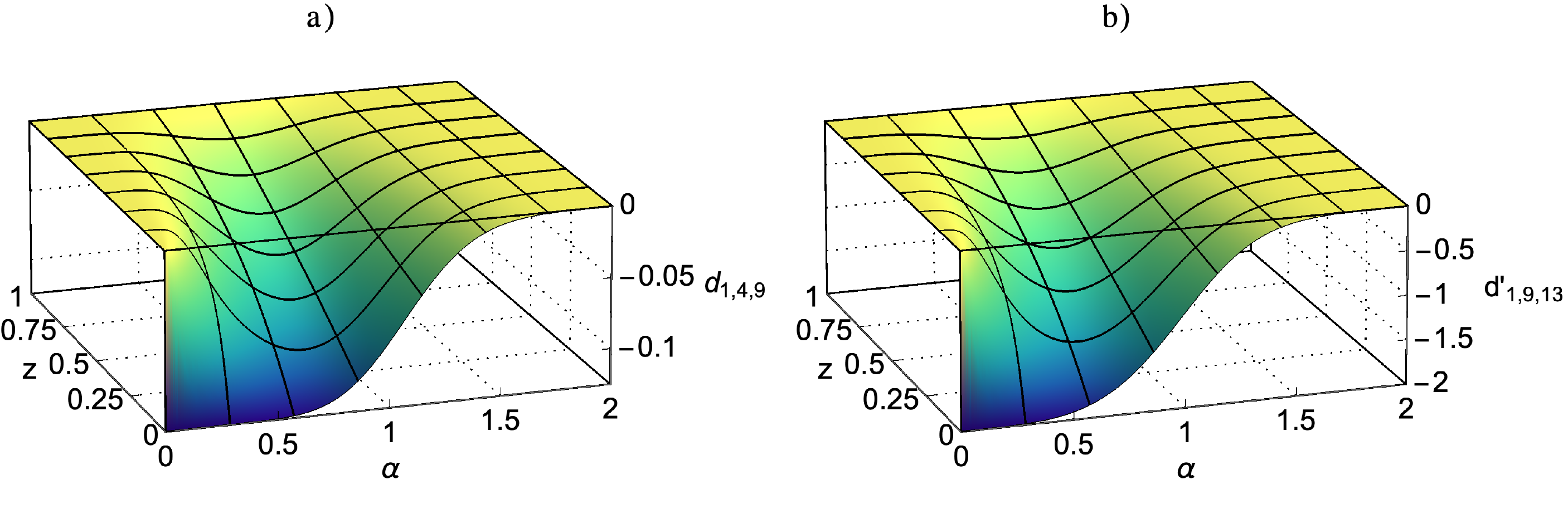}
	\caption{Comparison of the witnesses $d_{1,4,9}$ and $d'_{1,9,13}$ for the mixed Schrödinger cat state \eqref{eq:CatState} when $\beta = \alpha$ in a) and b), respectively.}
	\label{fig:d149d1939CatStates}
\end{figure*}
    \begin{equation}
        \begin{split}
            \boldsymbol D_{1,9,13} = \frac{1}{\abs{P_{123}}} \sum_{\sigma \in P_{123}} \Bigg\{&- \left(\boldsymbol L^{x2}_{a_{\sigma(1) \sigma(2)}}-\boldsymbol L^{y2}_{a_{\sigma(1) \sigma(2)}} \right)\left( \boldsymbol L^{x2}_{b_{\sigma(1) \sigma(2)}}-\boldsymbol L^{y2}_{b_{\sigma(1) \sigma(2)}}\right) \\
            &- \left\{\boldsymbol L^{x}_{a_{\sigma(1) \sigma(2)}},\boldsymbol L^{y}_{a_{\sigma(1) \sigma(2)}}\right\}\left\{\boldsymbol L^{x}_{b_{\sigma(1) \sigma(2)}},\boldsymbol L^{y}_{b_{\sigma(1) \sigma(2)}}\right\} \\
            &+ 2 \left[\left(\boldsymbol L^{x}_{a_{\sigma(1) \sigma(2)}} \boldsymbol L^{x}_{a_{\sigma(3) \sigma(1)}} +\boldsymbol L^{y}_{a_{\sigma(1) \sigma(2)}} \boldsymbol L^{y}_{a_{\sigma(3) \sigma(1)}}\right)\left(\boldsymbol L^{x}_{b_{\sigma(1) \sigma(2)}} \boldsymbol L^{x}_{b_{\sigma(3) \sigma(1)}} +\boldsymbol L^{y}_{b_{\sigma(1) \sigma(2)}} \boldsymbol L^{y}_{b_{\sigma(3) \sigma(1)}}\right)\right]\\
            &+ 2\left[\left(\boldsymbol L^{y}_{a_{\sigma(1) \sigma(2)}} \boldsymbol L^{x}_{a_{\sigma(3) \sigma(1)}} -\boldsymbol L^{x}_{a_{\sigma(1) \sigma(2)}} \boldsymbol L^{y}_{a_{\sigma(3) \sigma(1)}}\right)\left(\boldsymbol L^{y}_{b_{\sigma(1) \sigma(2)}} \boldsymbol L^{x}_{b_{\sigma(3) \sigma(1)}} -\boldsymbol L^{x}_{b_{\sigma(1) \sigma(2)}} \boldsymbol L^{y}_{b_{\sigma(3) \sigma(1)}}\right)\right]\\
            &-2 \left[ \boldsymbol n_{a_{\sigma(1)}}\boldsymbol n_{b_{\sigma(1)}}\left(\boldsymbol L^{x}_{a_{\sigma(2) \sigma(3)}} \boldsymbol L^{x}_{b_{\sigma(2) \sigma(3)}} +\boldsymbol L^{y}_{a_{\sigma(2) \sigma(3)}} \boldsymbol L^{y}_{b_{\sigma(2) \sigma(3)}}\right)\right]\\
            &+ \left(\boldsymbol n_{a_{\sigma(1)}}\boldsymbol n_{a_{\sigma(2)}}\boldsymbol n_{b_{\sigma(1)}}\boldsymbol n_{b_{\sigma(2)}}+\boldsymbol n_{a_{\sigma(1)}}\boldsymbol n_{b_{\sigma(1)}}\right)\\
            &+\frac{1}{2} \left(\boldsymbol n_{a_{\sigma(1)}}\boldsymbol n_{a_{\sigma(2)}}\boldsymbol n_{b_{\sigma(1)}}+\boldsymbol n_{a_{\sigma(1)}}\boldsymbol n_{a_{\sigma(3)}}\boldsymbol n_{b_{\sigma(1)}}+\boldsymbol n_{a_{\sigma(1)}}\boldsymbol n_{b_{\sigma(1)}}\boldsymbol n_{b_{\sigma(2)}}+\boldsymbol n_{a_{\sigma(1)}}\boldsymbol n_{b_{\sigma(1)}}\boldsymbol n_{b_{\sigma(3)}}\right)\\
            &-\frac{1}{2} \left[\left( \boldsymbol n_{a_{\sigma(3)}}+\boldsymbol n_{b_{\sigma(3)}}+1\right)\left( \left(\boldsymbol L^{x}_{a_{\sigma(1) \sigma(2)}} -i \boldsymbol L^{y}_{a_{\sigma(1) \sigma(2)}} \right) + \left( \boldsymbol L^{x}_{b_{\sigma(1) \sigma(2)}} +i \boldsymbol L^{y}_{b_{\sigma(1) \sigma(2)}}\right)\right) \right]\\
            &-\frac{1}{2} \left[\left( \boldsymbol n_{a_{\sigma(3)}}+\boldsymbol n_{b_{\sigma(3)}}+1\right) \left( \left(\boldsymbol L^{x}_{a_{\sigma(1) \sigma(2)}} +i \boldsymbol L^{y}_{a_{\sigma(1) \sigma(2)}} \right) + \left( \boldsymbol L^{x}_{b_{\sigma(1) \sigma(2)}} -i \boldsymbol L^{y}_{b_{\sigma(1) \sigma(2)}}\right)\right) \right] \Bigg\}.
        \end{split}
    \end{equation}
\end{widetext}
However, this expression is \textit{not} directly measurable due to all terms of the form
\begin{equation}
    \boldsymbol L^{x}_{a_{\sigma(1) \sigma(2)}} \boldsymbol L^{x}_{a_{\sigma(3) \sigma(1)}},
\end{equation}
which involve one and the same mode for several different spin operators. Passive interferometers can only simplify one of the two spin operators while complicating the other, thereby hindering the measurement of such observables when restricting to photon number measurements.

\section{Application of $d'_{1,9,13}$ to mixed Schrödinger cat states}
\label{app:d1913CatStates}
When applying $d'_{1,9,13}$ to the family of mixed Schrödinger cat states \eqref{eq:CatState}, we only need to evaluate three additional moments, which can be computed along the lines of \hyperref[App:valueCat]{Appendix} \ref{App:valueCat}. We find the expressions
\begin{equation}
    \begin{split}
        &\braket{\boldsymbol{a}^{\dagger} \boldsymbol{a}} = 2 \abs{\alpha}^2 N(\alpha,\beta,z) (1+(1-z)e^{-2\abs{\alpha}^2-2\abs{\beta}^2}),\\
        &\braket{\boldsymbol{a}^2 \boldsymbol{b}^{\dagger2}} = \alpha^2 \beta^{*2}, \\
        &\braket{\boldsymbol{a}^{\dagger2} \boldsymbol{b}^{2}} = \alpha^{*2} \beta^{2}.
    \end{split}
\end{equation}
We show the witness $d'_{1,9,13}$ for $\beta = \alpha$ in \autoref{fig:d149d1939CatStates}b and compare it to the witness $d_{1,4,9}$ in \autoref{fig:d149d1939CatStates}a. The resulting manifolds are surprisingly similar (up to an overall scaling factor). We observe that the witnesses work best around $\alpha \approx 1/2$, \textit{i.e.} when the two underlying coherent states partially overlap, and approach zero from below exponentially for $\alpha > 2$.

\end{appendix}

\bibliography{references.bib}

\begin{thebibliography}{49}%
\makeatletter
\providecommand \@ifxundefined [1]{%
 \@ifx{#1\undefined}
}%
\providecommand \@ifnum [1]{%
 \ifnum #1\expandafter \@firstoftwo
 \else \expandafter \@secondoftwo
 \fi
}%
\providecommand \@ifx [1]{%
 \ifx #1\expandafter \@firstoftwo
 \else \expandafter \@secondoftwo
 \fi
}%
\providecommand \natexlab [1]{#1}%
\providecommand \enquote  [1]{``#1''}%
\providecommand \bibnamefont  [1]{#1}%
\providecommand \bibfnamefont [1]{#1}%
\providecommand \citenamefont [1]{#1}%
\providecommand \href@noop [0]{\@secondoftwo}%
\providecommand \href [0]{\begingroup \@sanitize@url \@href}%
\providecommand \@href[1]{\@@startlink{#1}\@@href}%
\providecommand \@@href[1]{\endgroup#1\@@endlink}%
\providecommand \@sanitize@url [0]{\catcode `\\12\catcode `\$12\catcode
  `\&12\catcode `\#12\catcode `\^12\catcode `\_12\catcode `\%12\relax}%
\providecommand \@@startlink[1]{}%
\providecommand \@@endlink[0]{}%
\providecommand \url  [0]{\begingroup\@sanitize@url \@url }%
\providecommand \@url [1]{\endgroup\@href {#1}{\urlprefix }}%
\providecommand \urlprefix  [0]{URL }%
\providecommand \Eprint [0]{\href }%
\providecommand \doibase [0]{https://doi.org/}%
\providecommand \selectlanguage [0]{\@gobble}%
\providecommand \bibinfo  [0]{\@secondoftwo}%
\providecommand \bibfield  [0]{\@secondoftwo}%
\providecommand \translation [1]{[#1]}%
\providecommand \BibitemOpen [0]{}%
\providecommand \bibitemStop [0]{}%
\providecommand \bibitemNoStop [0]{.\EOS\space}%
\providecommand \EOS [0]{\spacefactor3000\relax}%
\providecommand \BibitemShut  [1]{\csname bibitem#1\endcsname}%
\let\auto@bib@innerbib\@empty
\bibitem [{\citenamefont {Plenio}\ and\ \citenamefont
  {Virmani}(2007)}]{Plenio2007}%
  \BibitemOpen
  \bibfield  {author} {\bibinfo {author} {\bibfnamefont {M.~B.}\ \bibnamefont
  {Plenio}}\ and\ \bibinfo {author} {\bibfnamefont {S.}~\bibnamefont
  {Virmani}},\ }\bibfield  {title} {\bibinfo {title} {{An Introduction to
  Entanglement Measures}},\ }\href {https://doi.org/10.5555/2011706.2011707}
  {\bibfield  {journal} {\bibinfo  {journal} {Quantum Inf. Comput.}\ }\textbf
  {\bibinfo {volume} {7}},\ \bibinfo {pages} {1–51} (\bibinfo {year}
  {2007})}\BibitemShut {NoStop}%
\bibitem [{\citenamefont {Horodecki}\ \emph {et~al.}(2009)\citenamefont
  {Horodecki}, \citenamefont {Horodecki}, \citenamefont {Horodecki},\ and\
  \citenamefont {Horodecki}}]{Horodecki2009}%
  \BibitemOpen
  \bibfield  {author} {\bibinfo {author} {\bibfnamefont {R.}~\bibnamefont
  {Horodecki}}, \bibinfo {author} {\bibfnamefont {P.}~\bibnamefont
  {Horodecki}}, \bibinfo {author} {\bibfnamefont {M.}~\bibnamefont
  {Horodecki}},\ and\ \bibinfo {author} {\bibfnamefont {K.}~\bibnamefont
  {Horodecki}},\ }\bibfield  {title} {\bibinfo {title} {{Quantum
  entanglement}},\ }\href {https://doi.org/10.1103/RevModPhys.81.865}
  {\bibfield  {journal} {\bibinfo  {journal} {Rev. Mod. Phys.}\ }\textbf
  {\bibinfo {volume} {81}},\ \bibinfo {pages} {865} (\bibinfo {year}
  {2009})}\BibitemShut {NoStop}%
\bibitem [{\citenamefont {G{\"u}hne}\ and\ \citenamefont
  {T{\'o}th}(2009)}]{Guehne2009}%
  \BibitemOpen
  \bibfield  {author} {\bibinfo {author} {\bibfnamefont {O.}~\bibnamefont
  {G{\"u}hne}}\ and\ \bibinfo {author} {\bibfnamefont {G.}~\bibnamefont
  {T{\'o}th}},\ }\bibfield  {title} {\bibinfo {title} {Entanglement
  detection},\ }\href {https://doi.org/10.1016/j.physrep.2009.02.004}
  {\bibfield  {journal} {\bibinfo  {journal} {Phys. Rep.}\ }\textbf {\bibinfo
  {volume} {474}},\ \bibinfo {pages} {1} (\bibinfo {year} {2009})}\BibitemShut
  {NoStop}%
\bibitem [{\citenamefont {Peres}(1996)}]{Peres1996}%
  \BibitemOpen
  \bibfield  {author} {\bibinfo {author} {\bibfnamefont {A.}~\bibnamefont
  {Peres}},\ }\bibfield  {title} {\bibinfo {title} {{Separability Criterion for
  Density Matrices}},\ }\href {https://doi.org/10.1103/PhysRevLett.77.1413}
  {\bibfield  {journal} {\bibinfo  {journal} {Phys. Rev. Lett.}\ }\textbf
  {\bibinfo {volume} {77}},\ \bibinfo {pages} {1413} (\bibinfo {year}
  {1996})}\BibitemShut {NoStop}%
\bibitem [{\citenamefont {Horodecki}\ \emph {et~al.}(1996)\citenamefont
  {Horodecki}, \citenamefont {Horodecki},\ and\ \citenamefont
  {Horodecki}}]{Horodecki1996}%
  \BibitemOpen
  \bibfield  {author} {\bibinfo {author} {\bibfnamefont {M.}~\bibnamefont
  {Horodecki}}, \bibinfo {author} {\bibfnamefont {P.}~\bibnamefont
  {Horodecki}},\ and\ \bibinfo {author} {\bibfnamefont {R.}~\bibnamefont
  {Horodecki}},\ }\bibfield  {title} {\bibinfo {title} {{Separability of mixed
  states: necessary and sufficient conditions}},\ }\href
  {https://doi.org/https://doi.org/10.1016/S0375-9601(96)00706-2} {\bibfield
  {journal} {\bibinfo  {journal} {Phys. Lett. A}\ }\textbf {\bibinfo {volume}
  {223}},\ \bibinfo {pages} {1} (\bibinfo {year} {1996})}\BibitemShut {NoStop}%
\bibitem [{\citenamefont {Braunstein}\ and\ \citenamefont {van
  Loock}(2005)}]{Braunstein2005b}%
  \BibitemOpen
  \bibfield  {author} {\bibinfo {author} {\bibfnamefont {S.~L.}\ \bibnamefont
  {Braunstein}}\ and\ \bibinfo {author} {\bibfnamefont {P.}~\bibnamefont {van
  Loock}},\ }\bibfield  {title} {\bibinfo {title} {{Quantum information with
  continuous variables}},\ }\href {https://doi.org/10.1103/RevModPhys.77.513}
  {\bibfield  {journal} {\bibinfo  {journal} {Rev. Mod. Phys.}\ }\textbf
  {\bibinfo {volume} {77}},\ \bibinfo {pages} {513} (\bibinfo {year}
  {2005})}\BibitemShut {NoStop}%
\bibitem [{\citenamefont {Weedbrook}\ \emph {et~al.}(2012)\citenamefont
  {Weedbrook}, \citenamefont {Pirandola}, \citenamefont
  {Garc\'{i}a-Patr\'{o}n}, \citenamefont {Cerf}, \citenamefont {Ralph},
  \citenamefont {Shapiro},\ and\ \citenamefont {Lloyd}}]{Weedbrook2012}%
  \BibitemOpen
  \bibfield  {author} {\bibinfo {author} {\bibfnamefont {C.}~\bibnamefont
  {Weedbrook}}, \bibinfo {author} {\bibfnamefont {S.}~\bibnamefont
  {Pirandola}}, \bibinfo {author} {\bibfnamefont {R.}~\bibnamefont
  {Garc\'{i}a-Patr\'{o}n}}, \bibinfo {author} {\bibfnamefont {N.~J.}\
  \bibnamefont {Cerf}}, \bibinfo {author} {\bibfnamefont {T.~C.}\ \bibnamefont
  {Ralph}}, \bibinfo {author} {\bibfnamefont {J.~H.}\ \bibnamefont {Shapiro}},\
  and\ \bibinfo {author} {\bibfnamefont {S.}~\bibnamefont {Lloyd}},\ }\bibfield
   {title} {\bibinfo {title} {{Gaussian quantum information}},\ }\href
  {https://doi.org/10.1103/RevModPhys.84.621} {\bibfield  {journal} {\bibinfo
  {journal} {Rev. Mod. Phys.}\ }\textbf {\bibinfo {volume} {84}},\ \bibinfo
  {pages} {621} (\bibinfo {year} {2012})}\BibitemShut {NoStop}%
\bibitem [{\citenamefont {Serafini}(2017)}]{Serafini2017}%
  \BibitemOpen
  \bibfield  {author} {\bibinfo {author} {\bibfnamefont {A.}~\bibnamefont
  {Serafini}},\ }\href {https://doi.org/10.1201/9781315118727} {\emph {\bibinfo
  {title} {{Quantum Continuous Variables}}}}\ (\bibinfo  {publisher} {CRC
  Press},\ \bibinfo {year} {2017})\BibitemShut {NoStop}%
\bibitem [{\citenamefont {Duan}\ \emph {et~al.}(2000)\citenamefont {Duan},
  \citenamefont {Giedke}, \citenamefont {Cirac},\ and\ \citenamefont
  {Zoller}}]{Duan2000}%
  \BibitemOpen
  \bibfield  {author} {\bibinfo {author} {\bibfnamefont {L.-M.}\ \bibnamefont
  {Duan}}, \bibinfo {author} {\bibfnamefont {G.}~\bibnamefont {Giedke}},
  \bibinfo {author} {\bibfnamefont {J.~I.}\ \bibnamefont {Cirac}},\ and\
  \bibinfo {author} {\bibfnamefont {P.}~\bibnamefont {Zoller}},\ }\bibfield
  {title} {\bibinfo {title} {{Inseparability Criterion for Continuous Variable
  Systems}},\ }\href {https://doi.org/10.1103/PhysRevLett.84.2722} {\bibfield
  {journal} {\bibinfo  {journal} {Phys. Rev. Lett.}\ }\textbf {\bibinfo
  {volume} {84}},\ \bibinfo {pages} {2722} (\bibinfo {year}
  {2000})}\BibitemShut {NoStop}%
\bibitem [{\citenamefont {Simon}(2000)}]{Simon2000}%
  \BibitemOpen
  \bibfield  {author} {\bibinfo {author} {\bibfnamefont {R.}~\bibnamefont
  {Simon}},\ }\bibfield  {title} {\bibinfo {title} {{Peres-Horodecki
  Separability Criterion for Continuous Variable Systems}},\ }\href
  {https://doi.org/10.1103/PhysRevLett.84.2726} {\bibfield  {journal} {\bibinfo
   {journal} {Phys. Rev. Lett.}\ }\textbf {\bibinfo {volume} {84}},\ \bibinfo
  {pages} {2726} (\bibinfo {year} {2000})}\BibitemShut {NoStop}%
\bibitem [{\citenamefont {Mancini}\ \emph {et~al.}(2002)\citenamefont
  {Mancini}, \citenamefont {Giovannetti}, \citenamefont {Vitali},\ and\
  \citenamefont {Tombesi}}]{Mancini2002}%
  \BibitemOpen
  \bibfield  {author} {\bibinfo {author} {\bibfnamefont {S.}~\bibnamefont
  {Mancini}}, \bibinfo {author} {\bibfnamefont {V.}~\bibnamefont
  {Giovannetti}}, \bibinfo {author} {\bibfnamefont {D.}~\bibnamefont
  {Vitali}},\ and\ \bibinfo {author} {\bibfnamefont {P.}~\bibnamefont
  {Tombesi}},\ }\bibfield  {title} {\bibinfo {title} {{Entangling Macroscopic
  Oscillators Exploiting Radiation Pressure}},\ }\href
  {https://doi.org/10.1103/PhysRevLett.88.120401} {\bibfield  {journal}
  {\bibinfo  {journal} {Phys. Rev. Lett.}\ }\textbf {\bibinfo {volume} {88}},\
  \bibinfo {pages} {120401} (\bibinfo {year} {2002})}\BibitemShut {NoStop}%
\bibitem [{\citenamefont {Giovannetti}\ \emph {et~al.}(2003)\citenamefont
  {Giovannetti}, \citenamefont {Mancini}, \citenamefont {Vitali},\ and\
  \citenamefont {Tombesi}}]{Giovannetti2003}%
  \BibitemOpen
  \bibfield  {author} {\bibinfo {author} {\bibfnamefont {V.}~\bibnamefont
  {Giovannetti}}, \bibinfo {author} {\bibfnamefont {S.}~\bibnamefont
  {Mancini}}, \bibinfo {author} {\bibfnamefont {D.}~\bibnamefont {Vitali}},\
  and\ \bibinfo {author} {\bibfnamefont {P.}~\bibnamefont {Tombesi}},\
  }\bibfield  {title} {\bibinfo {title} {{Characterizing the entanglement of
  bipartite quantum systems}},\ }\href
  {https://doi.org/10.1103/PhysRevA.67.022320} {\bibfield  {journal} {\bibinfo
  {journal} {Phys. Rev. A}\ }\textbf {\bibinfo {volume} {67}},\ \bibinfo
  {pages} {022320} (\bibinfo {year} {2003})}\BibitemShut {NoStop}%
\bibitem [{\citenamefont {Hertz}\ \emph {et~al.}(2016)\citenamefont {Hertz},
  \citenamefont {Karpov}, \citenamefont {Mandilara},\ and\ \citenamefont
  {Cerf}}]{Hertz2016}%
  \BibitemOpen
  \bibfield  {author} {\bibinfo {author} {\bibfnamefont {A.}~\bibnamefont
  {Hertz}}, \bibinfo {author} {\bibfnamefont {E.}~\bibnamefont {Karpov}},
  \bibinfo {author} {\bibfnamefont {A.}~\bibnamefont {Mandilara}},\ and\
  \bibinfo {author} {\bibfnamefont {N.~J.}\ \bibnamefont {Cerf}},\ }\bibfield
  {title} {\bibinfo {title} {{Detection of non-Gaussian entangled states with
  an improved continuous-variable separability criterion}},\ }\href
  {https://doi.org/10.1103/PhysRevA.93.032330} {\bibfield  {journal} {\bibinfo
  {journal} {Phys. Rev. A}\ }\textbf {\bibinfo {volume} {93}},\ \bibinfo
  {pages} {032330} (\bibinfo {year} {2016})}\BibitemShut {NoStop}%
\bibitem [{\citenamefont {Agarwal}\ and\ \citenamefont
  {Biswas}(2005)}]{Agarwal2005}%
  \BibitemOpen
  \bibfield  {author} {\bibinfo {author} {\bibfnamefont {G.~S.}\ \bibnamefont
  {Agarwal}}\ and\ \bibinfo {author} {\bibfnamefont {A.}~\bibnamefont
  {Biswas}},\ }\bibfield  {title} {\bibinfo {title} {{Inseparability
  inequalities for higher order moments for bipartite systems}},\ }\href
  {https://doi.org/10.1088/1367-2630/7/1/211} {\bibfield  {journal} {\bibinfo
  {journal} {New J. Phys.}\ }\textbf {\bibinfo {volume} {7}},\ \bibinfo {pages}
  {211} (\bibinfo {year} {2005})}\BibitemShut {NoStop}%
\bibitem [{\citenamefont {Walborn}\ \emph {et~al.}(2009)\citenamefont
  {Walborn}, \citenamefont {Taketani}, \citenamefont {Salles}, \citenamefont
  {Toscano},\ and\ \citenamefont {de~Matos~Filho}}]{Walborn2009}%
  \BibitemOpen
  \bibfield  {author} {\bibinfo {author} {\bibfnamefont {S.~P.}\ \bibnamefont
  {Walborn}}, \bibinfo {author} {\bibfnamefont {B.~G.}\ \bibnamefont
  {Taketani}}, \bibinfo {author} {\bibfnamefont {A.}~\bibnamefont {Salles}},
  \bibinfo {author} {\bibfnamefont {F.}~\bibnamefont {Toscano}},\ and\ \bibinfo
  {author} {\bibfnamefont {R.~L.}\ \bibnamefont {de~Matos~Filho}},\ }\bibfield
  {title} {\bibinfo {title} {{Entropic Entanglement Criteria for Continuous
  Variables}},\ }\href {https://doi.org/10.1103/PhysRevLett.103.160505}
  {\bibfield  {journal} {\bibinfo  {journal} {Phys. Rev. Lett.}\ }\textbf
  {\bibinfo {volume} {103}},\ \bibinfo {pages} {160505} (\bibinfo {year}
  {2009})}\BibitemShut {NoStop}%
\bibitem [{\citenamefont {Walborn}\ \emph {et~al.}(2011)\citenamefont
  {Walborn}, \citenamefont {Salles}, \citenamefont {Gomes}, \citenamefont
  {Toscano},\ and\ \citenamefont {Ribeiro}}]{Walborn2011}%
  \BibitemOpen
  \bibfield  {author} {\bibinfo {author} {\bibfnamefont {S.~P.}\ \bibnamefont
  {Walborn}}, \bibinfo {author} {\bibfnamefont {A.}~\bibnamefont {Salles}},
  \bibinfo {author} {\bibfnamefont {R.~M.}\ \bibnamefont {Gomes}}, \bibinfo
  {author} {\bibfnamefont {F.}~\bibnamefont {Toscano}},\ and\ \bibinfo {author}
  {\bibfnamefont {P.~H.~S.}\ \bibnamefont {Ribeiro}},\ }\bibfield  {title}
  {\bibinfo {title} {{Revealing Hidden Einstein-Podolsky-Rosen Nonlocality}},\
  }\href {https://doi.org/10.1103/PhysRevLett.106.130402} {\bibfield  {journal}
  {\bibinfo  {journal} {Phys. Rev. Lett.}\ }\textbf {\bibinfo {volume} {106}},\
  \bibinfo {pages} {130402} (\bibinfo {year} {2011})}\BibitemShut {NoStop}%
\bibitem [{\citenamefont {Saboia}\ \emph {et~al.}(2011)\citenamefont {Saboia},
  \citenamefont {Toscano},\ and\ \citenamefont {Walborn}}]{Saboia2011}%
  \BibitemOpen
  \bibfield  {author} {\bibinfo {author} {\bibfnamefont {A.}~\bibnamefont
  {Saboia}}, \bibinfo {author} {\bibfnamefont {F.}~\bibnamefont {Toscano}},\
  and\ \bibinfo {author} {\bibfnamefont {S.~P.}\ \bibnamefont {Walborn}},\
  }\bibfield  {title} {\bibinfo {title} {{Family of continuous-variable
  entanglement criteria using general entropy functions}},\ }\href
  {https://doi.org/10.1103/PhysRevA.83.032307} {\bibfield  {journal} {\bibinfo
  {journal} {Phys. Rev. A}\ }\textbf {\bibinfo {volume} {83}},\ \bibinfo
  {pages} {032307} (\bibinfo {year} {2011})}\BibitemShut {NoStop}%
\bibitem [{\citenamefont {Tasca}\ \emph {et~al.}(2013)\citenamefont {Tasca},
  \citenamefont {Rudnicki}, \citenamefont {Gomes}, \citenamefont {Toscano},\
  and\ \citenamefont {Walborn}}]{Tasca2013}%
  \BibitemOpen
  \bibfield  {author} {\bibinfo {author} {\bibfnamefont {D.~S.}\ \bibnamefont
  {Tasca}}, \bibinfo {author} {\bibfnamefont {{\L}.}~\bibnamefont {Rudnicki}},
  \bibinfo {author} {\bibfnamefont {R.~M.}\ \bibnamefont {Gomes}}, \bibinfo
  {author} {\bibfnamefont {F.}~\bibnamefont {Toscano}},\ and\ \bibinfo {author}
  {\bibfnamefont {S.~P.}\ \bibnamefont {Walborn}},\ }\bibfield  {title}
  {\bibinfo {title} {{Reliable Entanglement Detection under Coarse-Grained
  Measurements}},\ }\href {https://doi.org/10.1103/PhysRevLett.110.210502}
  {\bibfield  {journal} {\bibinfo  {journal} {Phys. Rev. Lett.}\ }\textbf
  {\bibinfo {volume} {110}},\ \bibinfo {pages} {210502} (\bibinfo {year}
  {2013})}\BibitemShut {NoStop}%
\bibitem [{\citenamefont {Schneeloch}\ and\ \citenamefont
  {Howland}(2018)}]{Schneeloch2018}%
  \BibitemOpen
  \bibfield  {author} {\bibinfo {author} {\bibfnamefont {J.}~\bibnamefont
  {Schneeloch}}\ and\ \bibinfo {author} {\bibfnamefont {G.~A.}\ \bibnamefont
  {Howland}},\ }\bibfield  {title} {\bibinfo {title} {{Quantifying
  high-dimensional entanglement with Einstein-Podolsky-Rosen correlations}},\
  }\href {https://doi.org/10.1103/PhysRevA.97.042338} {\bibfield  {journal}
  {\bibinfo  {journal} {Phys. Rev. A}\ }\textbf {\bibinfo {volume} {97}},\
  \bibinfo {pages} {042338} (\bibinfo {year} {2018})}\BibitemShut {NoStop}%
\bibitem [{\citenamefont {Floerchinger}\ \emph {et~al.}(2021)\citenamefont
  {Floerchinger}, \citenamefont {Haas},\ and\ \citenamefont
  {M\"uller-Groeling}}]{Haas2021}%
  \BibitemOpen
  \bibfield  {author} {\bibinfo {author} {\bibfnamefont {S.}~\bibnamefont
  {Floerchinger}}, \bibinfo {author} {\bibfnamefont {T.}~\bibnamefont {Haas}},\
  and\ \bibinfo {author} {\bibfnamefont {H.}~\bibnamefont
  {M\"uller-Groeling}},\ }\bibfield  {title} {\bibinfo {title} {{Wehrl entropy,
  entropic uncertainty relations, and entanglement}},\ }\href
  {https://doi.org/10.1103/PhysRevA.103.062222} {\bibfield  {journal} {\bibinfo
   {journal} {Phys. Rev. A}\ }\textbf {\bibinfo {volume} {103}},\ \bibinfo
  {pages} {062222} (\bibinfo {year} {2021})}\BibitemShut {NoStop}%
\bibitem [{\citenamefont {Floerchinger}\ \emph {et~al.}(2022)\citenamefont
  {Floerchinger}, \citenamefont {G\"arttner}, \citenamefont {Haas},\ and\
  \citenamefont {Stockdale}}]{Haas2022a}%
  \BibitemOpen
  \bibfield  {author} {\bibinfo {author} {\bibfnamefont {S.}~\bibnamefont
  {Floerchinger}}, \bibinfo {author} {\bibfnamefont {M.}~\bibnamefont
  {G\"arttner}}, \bibinfo {author} {\bibfnamefont {T.}~\bibnamefont {Haas}},\
  and\ \bibinfo {author} {\bibfnamefont {O.~R.}\ \bibnamefont {Stockdale}},\
  }\bibfield  {title} {\bibinfo {title} {{Entropic entanglement criteria in
  phase space}},\ }\href {https://doi.org/10.1103/PhysRevA.105.012409}
  {\bibfield  {journal} {\bibinfo  {journal} {Phys. Rev. A}\ }\textbf {\bibinfo
  {volume} {105}},\ \bibinfo {pages} {012409} (\bibinfo {year}
  {2022})}\BibitemShut {NoStop}%
\bibitem [{\citenamefont {Gärttner}\ \emph
  {et~al.}(2022{\natexlab{a}})\citenamefont {Gärttner}, \citenamefont {Haas},\
  and\ \citenamefont {Noll}}]{Haas2022b}%
  \BibitemOpen
  \bibfield  {author} {\bibinfo {author} {\bibfnamefont {M.}~\bibnamefont
  {Gärttner}}, \bibinfo {author} {\bibfnamefont {T.}~\bibnamefont {Haas}},\
  and\ \bibinfo {author} {\bibfnamefont {J.}~\bibnamefont {Noll}},\ }\bibfield
  {title} {\bibinfo {title} {{General class of continuous variable entanglement
  criteria}},\ }\href@noop {} {\bibfield  {journal} {\bibinfo  {journal}
  {\href{https://arxiv.org/abs/2211.17160}{arXiv:2211.17160}}\ } (\bibinfo
  {year} {2022}{\natexlab{a}})}\BibitemShut {NoStop}%
\bibitem [{\citenamefont {Gärttner}\ \emph
  {et~al.}(2022{\natexlab{b}})\citenamefont {Gärttner}, \citenamefont {Haas},\
  and\ \citenamefont {Noll}}]{Haas2022c}%
  \BibitemOpen
  \bibfield  {author} {\bibinfo {author} {\bibfnamefont {M.}~\bibnamefont
  {Gärttner}}, \bibinfo {author} {\bibfnamefont {T.}~\bibnamefont {Haas}},\
  and\ \bibinfo {author} {\bibfnamefont {J.}~\bibnamefont {Noll}},\ }\bibfield
  {title} {\bibinfo {title} {{Detecting continuous variable entanglement in
  phase space with the Q-distribution}},\ }\href@noop {} {\bibfield  {journal}
  {\bibinfo  {journal}
  {\href{https://arxiv.org/abs/2211.17165}{arXiv:2211.17165}}\ } (\bibinfo
  {year} {2022}{\natexlab{b}})}\BibitemShut {NoStop}%
\bibitem [{\citenamefont {Gärttner}\ \emph
  {et~al.}(2022{\natexlab{c}})\citenamefont {Gärttner}, \citenamefont {Haas},\
  and\ \citenamefont {Noll}}]{Haas2022d}%
  \BibitemOpen
  \bibfield  {author} {\bibinfo {author} {\bibfnamefont {M.}~\bibnamefont
  {Gärttner}}, \bibinfo {author} {\bibfnamefont {T.}~\bibnamefont {Haas}},\
  and\ \bibinfo {author} {\bibfnamefont {J.}~\bibnamefont {Noll}},\ }\bibfield
  {title} {\bibinfo {title} {{Optimizing detection of continuous variable
  entanglement for limited data}},\ }\href@noop {} {\bibfield  {journal}
  {\bibinfo  {journal}
  {\href{https://arxiv.org/abs/2211.17168}{arXiv:2211.17168}}\ } (\bibinfo
  {year} {2022}{\natexlab{c}})}\BibitemShut {NoStop}%
\bibitem [{\citenamefont {Dong}\ \emph {et~al.}(2008)\citenamefont {Dong},
  \citenamefont {Lassen}, \citenamefont {Heersink}, \citenamefont {Marquardt},
  \citenamefont {Filip}, \citenamefont {Leuchs},\ and\ \citenamefont
  {Andersen}}]{Dong2008}%
  \BibitemOpen
  \bibfield  {author} {\bibinfo {author} {\bibfnamefont {R.}~\bibnamefont
  {Dong}}, \bibinfo {author} {\bibfnamefont {M.}~\bibnamefont {Lassen}},
  \bibinfo {author} {\bibfnamefont {J.}~\bibnamefont {Heersink}}, \bibinfo
  {author} {\bibfnamefont {C.}~\bibnamefont {Marquardt}}, \bibinfo {author}
  {\bibfnamefont {R.}~\bibnamefont {Filip}}, \bibinfo {author} {\bibfnamefont
  {G.}~\bibnamefont {Leuchs}},\ and\ \bibinfo {author} {\bibfnamefont {U.~L.}\
  \bibnamefont {Andersen}},\ }\bibfield  {title} {\bibinfo {title}
  {{Experimental entanglement distillation of mesoscopic quantum states}},\
  }\href {https://doi.org/10.1038/nphys1112} {\bibfield  {journal} {\bibinfo
  {journal} {Nat. Phys.}\ }\textbf {\bibinfo {volume} {4}},\ \bibinfo {pages}
  {919} (\bibinfo {year} {2008})}\BibitemShut {NoStop}%
\bibitem [{\citenamefont {Schneeloch}\ \emph {et~al.}(2019)\citenamefont
  {Schneeloch}, \citenamefont {Tison}, \citenamefont {Fanto}, \citenamefont
  {Alsing},\ and\ \citenamefont {Howland}}]{Schneeloch2019}%
  \BibitemOpen
  \bibfield  {author} {\bibinfo {author} {\bibfnamefont {J.}~\bibnamefont
  {Schneeloch}}, \bibinfo {author} {\bibfnamefont {C.~C.}\ \bibnamefont
  {Tison}}, \bibinfo {author} {\bibfnamefont {M.~L.}\ \bibnamefont {Fanto}},
  \bibinfo {author} {\bibfnamefont {P.~M.}\ \bibnamefont {Alsing}},\ and\
  \bibinfo {author} {\bibfnamefont {G.~A.}\ \bibnamefont {Howland}},\
  }\bibfield  {title} {\bibinfo {title} {{Quantifying entanglement in a
  68-billion-dimensional quantum state space}},\ }\href
  {https://doi.org/10.1038/s41467-019-10810-z} {\bibfield  {journal} {\bibinfo
  {journal} {Nat. Commun.}\ }\textbf {\bibinfo {volume} {10}},\ \bibinfo
  {pages} {1} (\bibinfo {year} {2019})}\BibitemShut {NoStop}%
\bibitem [{\citenamefont {Asavanant}\ \emph {et~al.}(2019)\citenamefont
  {Asavanant}, \citenamefont {Shiozawa}, \citenamefont {Yokoyama},
  \citenamefont {Charoensombutamon}, \citenamefont {Emura}, \citenamefont
  {Alexander}, \citenamefont {Takeda}, \citenamefont {Yoshikawa}, \citenamefont
  {Menicucci}, \citenamefont {Yonezawa},\ and\ \citenamefont
  {Furusawa}}]{Asavanant2019}%
  \BibitemOpen
  \bibfield  {author} {\bibinfo {author} {\bibfnamefont {W.}~\bibnamefont
  {Asavanant}}, \bibinfo {author} {\bibfnamefont {Y.}~\bibnamefont {Shiozawa}},
  \bibinfo {author} {\bibfnamefont {S.}~\bibnamefont {Yokoyama}}, \bibinfo
  {author} {\bibfnamefont {B.}~\bibnamefont {Charoensombutamon}}, \bibinfo
  {author} {\bibfnamefont {H.}~\bibnamefont {Emura}}, \bibinfo {author}
  {\bibfnamefont {R.}~\bibnamefont {Alexander}}, \bibinfo {author}
  {\bibfnamefont {S.}~\bibnamefont {Takeda}}, \bibinfo {author} {\bibfnamefont
  {J.-i.}\ \bibnamefont {Yoshikawa}}, \bibinfo {author} {\bibfnamefont
  {N.}~\bibnamefont {Menicucci}}, \bibinfo {author} {\bibfnamefont
  {H.}~\bibnamefont {Yonezawa}},\ and\ \bibinfo {author} {\bibfnamefont
  {A.}~\bibnamefont {Furusawa}},\ }\bibfield  {title} {\bibinfo {title}
  {{Generation of time-domain-multiplexed two-dimensional cluster state}},\
  }\href {https://doi.org/10.1126/science.aay2645} {\bibfield  {journal}
  {\bibinfo  {journal} {Science}\ }\textbf {\bibinfo {volume} {366}},\ \bibinfo
  {pages} {373} (\bibinfo {year} {2019})}\BibitemShut {NoStop}%
\bibitem [{\citenamefont {Qin}\ \emph {et~al.}(2019)\citenamefont {Qin},
  \citenamefont {Gessner}, \citenamefont {Ren}, \citenamefont {Deng},
  \citenamefont {Han}, \citenamefont {Li}, \citenamefont {Su}, \citenamefont
  {Smerzi},\ and\ \citenamefont {Peng}}]{Qin2019}%
  \BibitemOpen
  \bibfield  {author} {\bibinfo {author} {\bibfnamefont {Z.}~\bibnamefont
  {Qin}}, \bibinfo {author} {\bibfnamefont {M.}~\bibnamefont {Gessner}},
  \bibinfo {author} {\bibfnamefont {Z.}~\bibnamefont {Ren}}, \bibinfo {author}
  {\bibfnamefont {X.}~\bibnamefont {Deng}}, \bibinfo {author} {\bibfnamefont
  {D.}~\bibnamefont {Han}}, \bibinfo {author} {\bibfnamefont {W.}~\bibnamefont
  {Li}}, \bibinfo {author} {\bibfnamefont {X.}~\bibnamefont {Su}}, \bibinfo
  {author} {\bibfnamefont {A.}~\bibnamefont {Smerzi}},\ and\ \bibinfo {author}
  {\bibfnamefont {K.}~\bibnamefont {Peng}},\ }\bibfield  {title} {\bibinfo
  {title} {{Characterizing the multipartite continuous-variable entanglement
  structure from squeezing coefficients and the Fisher information}},\ }\href
  {https://doi.org/10.1038/s41534-018-0119-6} {\bibfield  {journal} {\bibinfo
  {journal} {npj Quantum Inf.}\ }\textbf {\bibinfo {volume} {5}},\ \bibinfo
  {pages} {3} (\bibinfo {year} {2019})}\BibitemShut {NoStop}%
\bibitem [{\citenamefont {Moody}\ \emph {et~al.}(2022)\citenamefont {Moody},
  \citenamefont {Sorger}, \citenamefont {Blumenthal}, \citenamefont
  {Juodawlkis}, \citenamefont {Loh}, \citenamefont {Sorace-Agaskar},
  \citenamefont {Jones}, \citenamefont {Balram}, \citenamefont {Matthews},
  \citenamefont {Laing}, \citenamefont {Davanco}, \citenamefont {Chang},
  \citenamefont {Bowers}, \citenamefont {Quack}, \citenamefont {Galland},
  \citenamefont {Aharonovich}, \citenamefont {Wolff}, \citenamefont {Schuck},
  \citenamefont {Sinclair}, \citenamefont {Lončar}, \citenamefont
  {Komljenovic}, \citenamefont {Weld}, \citenamefont {Mookherjea},
  \citenamefont {Buckley}, \citenamefont {Radulaski}, \citenamefont
  {Reitzenstein}, \citenamefont {Pingault}, \citenamefont {Machielse},
  \citenamefont {Mukhopadhyay}, \citenamefont {Akimov}, \citenamefont
  {Zheltikov}, \citenamefont {Agarwal}, \citenamefont {Srinivasan},
  \citenamefont {Lu}, \citenamefont {Tang}, \citenamefont {Jiang},
  \citenamefont {McKenna}, \citenamefont {Safavi-Naeini}, \citenamefont
  {Steinhauer}, \citenamefont {Elshaari}, \citenamefont {Zwiller},
  \citenamefont {Davids}, \citenamefont {Martinez}, \citenamefont {Gehl},
  \citenamefont {Chiaverini}, \citenamefont {Mehta}, \citenamefont {Romero},
  \citenamefont {Lingaraju}, \citenamefont {Weiner}, \citenamefont {Peace},
  \citenamefont {Cernansky}, \citenamefont {Lobino}, \citenamefont {Diamanti},
  \citenamefont {Vidarte},\ and\ \citenamefont {Camacho}}]{Moody2022}%
  \BibitemOpen
  \bibfield  {author} {\bibinfo {author} {\bibfnamefont {G.}~\bibnamefont
  {Moody}}, \bibinfo {author} {\bibfnamefont {V.~J.}\ \bibnamefont {Sorger}},
  \bibinfo {author} {\bibfnamefont {D.~J.}\ \bibnamefont {Blumenthal}},
  \bibinfo {author} {\bibfnamefont {P.~W.}\ \bibnamefont {Juodawlkis}},
  \bibinfo {author} {\bibfnamefont {W.}~\bibnamefont {Loh}}, \bibinfo {author}
  {\bibfnamefont {C.}~\bibnamefont {Sorace-Agaskar}}, \bibinfo {author}
  {\bibfnamefont {A.~E.}\ \bibnamefont {Jones}}, \bibinfo {author}
  {\bibfnamefont {K.~C.}\ \bibnamefont {Balram}}, \bibinfo {author}
  {\bibfnamefont {J.~C.~F.}\ \bibnamefont {Matthews}}, \bibinfo {author}
  {\bibfnamefont {A.}~\bibnamefont {Laing}}, \bibinfo {author} {\bibfnamefont
  {M.}~\bibnamefont {Davanco}}, \bibinfo {author} {\bibfnamefont
  {L.}~\bibnamefont {Chang}}, \bibinfo {author} {\bibfnamefont {J.~E.}\
  \bibnamefont {Bowers}}, \bibinfo {author} {\bibfnamefont {N.}~\bibnamefont
  {Quack}}, \bibinfo {author} {\bibfnamefont {C.}~\bibnamefont {Galland}},
  \bibinfo {author} {\bibfnamefont {I.}~\bibnamefont {Aharonovich}}, \bibinfo
  {author} {\bibfnamefont {M.~A.}\ \bibnamefont {Wolff}}, \bibinfo {author}
  {\bibfnamefont {C.}~\bibnamefont {Schuck}}, \bibinfo {author} {\bibfnamefont
  {N.}~\bibnamefont {Sinclair}}, \bibinfo {author} {\bibfnamefont
  {M.}~\bibnamefont {Lončar}}, \bibinfo {author} {\bibfnamefont
  {T.}~\bibnamefont {Komljenovic}}, \bibinfo {author} {\bibfnamefont
  {D.}~\bibnamefont {Weld}}, \bibinfo {author} {\bibfnamefont {S.}~\bibnamefont
  {Mookherjea}}, \bibinfo {author} {\bibfnamefont {S.}~\bibnamefont {Buckley}},
  \bibinfo {author} {\bibfnamefont {M.}~\bibnamefont {Radulaski}}, \bibinfo
  {author} {\bibfnamefont {S.}~\bibnamefont {Reitzenstein}}, \bibinfo {author}
  {\bibfnamefont {B.}~\bibnamefont {Pingault}}, \bibinfo {author}
  {\bibfnamefont {B.}~\bibnamefont {Machielse}}, \bibinfo {author}
  {\bibfnamefont {D.}~\bibnamefont {Mukhopadhyay}}, \bibinfo {author}
  {\bibfnamefont {A.}~\bibnamefont {Akimov}}, \bibinfo {author} {\bibfnamefont
  {A.}~\bibnamefont {Zheltikov}}, \bibinfo {author} {\bibfnamefont {G.~S.}\
  \bibnamefont {Agarwal}}, \bibinfo {author} {\bibfnamefont {K.}~\bibnamefont
  {Srinivasan}}, \bibinfo {author} {\bibfnamefont {J.}~\bibnamefont {Lu}},
  \bibinfo {author} {\bibfnamefont {H.~X.}\ \bibnamefont {Tang}}, \bibinfo
  {author} {\bibfnamefont {W.}~\bibnamefont {Jiang}}, \bibinfo {author}
  {\bibfnamefont {T.~P.}\ \bibnamefont {McKenna}}, \bibinfo {author}
  {\bibfnamefont {A.~H.}\ \bibnamefont {Safavi-Naeini}}, \bibinfo {author}
  {\bibfnamefont {S.}~\bibnamefont {Steinhauer}}, \bibinfo {author}
  {\bibfnamefont {A.~W.}\ \bibnamefont {Elshaari}}, \bibinfo {author}
  {\bibfnamefont {V.}~\bibnamefont {Zwiller}}, \bibinfo {author} {\bibfnamefont
  {P.~S.}\ \bibnamefont {Davids}}, \bibinfo {author} {\bibfnamefont
  {N.}~\bibnamefont {Martinez}}, \bibinfo {author} {\bibfnamefont
  {M.}~\bibnamefont {Gehl}}, \bibinfo {author} {\bibfnamefont {J.}~\bibnamefont
  {Chiaverini}}, \bibinfo {author} {\bibfnamefont {K.~K.}\ \bibnamefont
  {Mehta}}, \bibinfo {author} {\bibfnamefont {J.}~\bibnamefont {Romero}},
  \bibinfo {author} {\bibfnamefont {N.~B.}\ \bibnamefont {Lingaraju}}, \bibinfo
  {author} {\bibfnamefont {A.~M.}\ \bibnamefont {Weiner}}, \bibinfo {author}
  {\bibfnamefont {D.}~\bibnamefont {Peace}}, \bibinfo {author} {\bibfnamefont
  {R.}~\bibnamefont {Cernansky}}, \bibinfo {author} {\bibfnamefont
  {M.}~\bibnamefont {Lobino}}, \bibinfo {author} {\bibfnamefont
  {E.}~\bibnamefont {Diamanti}}, \bibinfo {author} {\bibfnamefont {L.~T.}\
  \bibnamefont {Vidarte}},\ and\ \bibinfo {author} {\bibfnamefont {R.~M.}\
  \bibnamefont {Camacho}},\ }\bibfield  {title} {\bibinfo {title} {{2022
  Roadmap on integrated quantum photonics}},\ }\href
  {https://doi.org/10.1088/2515-7647/ac1ef4} {\bibfield  {journal} {\bibinfo
  {journal} {J. Phys. Photonics}\ }\textbf {\bibinfo {volume} {4}},\ \bibinfo
  {pages} {012501} (\bibinfo {year} {2022})}\BibitemShut {NoStop}%
\bibitem [{\citenamefont {Gross}\ \emph {et~al.}(2011)\citenamefont {Gross},
  \citenamefont {Strobel}, \citenamefont {Nicklas}, \citenamefont {Zibold},
  \citenamefont {Bar-Gill}, \citenamefont {Kurizki},\ and\ \citenamefont
  {Oberthaler}}]{Gross2011}%
  \BibitemOpen
  \bibfield  {author} {\bibinfo {author} {\bibfnamefont {C.}~\bibnamefont
  {Gross}}, \bibinfo {author} {\bibfnamefont {H.}~\bibnamefont {Strobel}},
  \bibinfo {author} {\bibfnamefont {E.}~\bibnamefont {Nicklas}}, \bibinfo
  {author} {\bibfnamefont {T.}~\bibnamefont {Zibold}}, \bibinfo {author}
  {\bibfnamefont {N.}~\bibnamefont {Bar-Gill}}, \bibinfo {author}
  {\bibfnamefont {G.}~\bibnamefont {Kurizki}},\ and\ \bibinfo {author}
  {\bibfnamefont {M.~K.}\ \bibnamefont {Oberthaler}},\ }\bibfield  {title}
  {\bibinfo {title} {{Atomic homodyne detection of continuous-variable
  entangled twin-atom states}},\ }\href {https://doi.org/10.1038/nature10654}
  {\bibfield  {journal} {\bibinfo  {journal} {Nature}\ }\textbf {\bibinfo
  {volume} {480}},\ \bibinfo {pages} {219} (\bibinfo {year}
  {2011})}\BibitemShut {NoStop}%
\bibitem [{\citenamefont {Strobel}\ \emph {et~al.}(2014)\citenamefont
  {Strobel}, \citenamefont {Muessel}, \citenamefont {Linnemann}, \citenamefont
  {Zibold}, \citenamefont {Hume}, \citenamefont {Pezz{\`e}}, \citenamefont
  {Smerzi},\ and\ \citenamefont {Oberthaler}}]{Strobel2014}%
  \BibitemOpen
  \bibfield  {author} {\bibinfo {author} {\bibfnamefont {H.}~\bibnamefont
  {Strobel}}, \bibinfo {author} {\bibfnamefont {W.}~\bibnamefont {Muessel}},
  \bibinfo {author} {\bibfnamefont {D.}~\bibnamefont {Linnemann}}, \bibinfo
  {author} {\bibfnamefont {T.}~\bibnamefont {Zibold}}, \bibinfo {author}
  {\bibfnamefont {D.~B.}\ \bibnamefont {Hume}}, \bibinfo {author}
  {\bibfnamefont {L.}~\bibnamefont {Pezz{\`e}}}, \bibinfo {author}
  {\bibfnamefont {A.}~\bibnamefont {Smerzi}},\ and\ \bibinfo {author}
  {\bibfnamefont {M.~K.}\ \bibnamefont {Oberthaler}},\ }\bibfield  {title}
  {\bibinfo {title} {{Fisher information and entanglement of non-Gaussian spin
  states}},\ }\href {https://doi.org/10.1126/science.1250147} {\bibfield
  {journal} {\bibinfo  {journal} {Science}\ }\textbf {\bibinfo {volume}
  {345}},\ \bibinfo {pages} {424} (\bibinfo {year} {2014})}\BibitemShut
  {NoStop}%
\bibitem [{\citenamefont {Peise}\ \emph {et~al.}(2015)\citenamefont {Peise},
  \citenamefont {Kruse}, \citenamefont {Lange}, \citenamefont {Lücke},
  \citenamefont {Pezzè}, \citenamefont {Arlt}, \citenamefont {Ertmer},
  \citenamefont {Hammerer}, \citenamefont {Santos}, \citenamefont {Smerzi},\
  and\ \citenamefont {Klempt}}]{Peise2015}%
  \BibitemOpen
  \bibfield  {author} {\bibinfo {author} {\bibfnamefont {J.}~\bibnamefont
  {Peise}}, \bibinfo {author} {\bibfnamefont {I.}~\bibnamefont {Kruse}},
  \bibinfo {author} {\bibfnamefont {K.}~\bibnamefont {Lange}}, \bibinfo
  {author} {\bibfnamefont {B.}~\bibnamefont {Lücke}}, \bibinfo {author}
  {\bibfnamefont {L.}~\bibnamefont {Pezzè}}, \bibinfo {author} {\bibfnamefont
  {J.}~\bibnamefont {Arlt}}, \bibinfo {author} {\bibfnamefont {W.}~\bibnamefont
  {Ertmer}}, \bibinfo {author} {\bibfnamefont {K.}~\bibnamefont {Hammerer}},
  \bibinfo {author} {\bibfnamefont {L.}~\bibnamefont {Santos}}, \bibinfo
  {author} {\bibfnamefont {A.}~\bibnamefont {Smerzi}},\ and\ \bibinfo {author}
  {\bibfnamefont {C.}~\bibnamefont {Klempt}},\ }\bibfield  {title} {\bibinfo
  {title} {{Satisfying the Einstein-Podolsky-Rosen criterion with massive
  particles}},\ }\href {https://doi.org/10.1038/ncomms9984} {\bibfield
  {journal} {\bibinfo  {journal} {Nat. Commun.}\ }\textbf {\bibinfo {volume}
  {6}},\ \bibinfo {pages} {8984} (\bibinfo {year} {2015})}\BibitemShut
  {NoStop}%
\bibitem [{\citenamefont {Fadel}\ \emph {et~al.}(2018)\citenamefont {Fadel},
  \citenamefont {Zibold}, \citenamefont {Décamps},\ and\ \citenamefont
  {Treutlein}}]{Fadel2018}%
  \BibitemOpen
  \bibfield  {author} {\bibinfo {author} {\bibfnamefont {M.}~\bibnamefont
  {Fadel}}, \bibinfo {author} {\bibfnamefont {T.}~\bibnamefont {Zibold}},
  \bibinfo {author} {\bibfnamefont {B.}~\bibnamefont {Décamps}},\ and\
  \bibinfo {author} {\bibfnamefont {P.}~\bibnamefont {Treutlein}},\ }\bibfield
  {title} {\bibinfo {title} {{Spatial entanglement patterns and
  Einstein-Podolsky-Rosen steering in Bose-Einstein condensates}},\ }\href
  {https://doi.org/10.1126/science.aao1850} {\bibfield  {journal} {\bibinfo
  {journal} {Science}\ }\textbf {\bibinfo {volume} {360}},\ \bibinfo {pages}
  {409} (\bibinfo {year} {2018})}\BibitemShut {NoStop}%
\bibitem [{\citenamefont {Kunkel}\ \emph {et~al.}(2018)\citenamefont {Kunkel},
  \citenamefont {Prüfer}, \citenamefont {Strobel}, \citenamefont {Linnemann},
  \citenamefont {Fr{\"o}lian}, \citenamefont {Gasenzer}, \citenamefont
  {Gärttner},\ and\ \citenamefont {Oberthaler}}]{Kunkel2018}%
  \BibitemOpen
  \bibfield  {author} {\bibinfo {author} {\bibfnamefont {P.}~\bibnamefont
  {Kunkel}}, \bibinfo {author} {\bibfnamefont {M.}~\bibnamefont {Prüfer}},
  \bibinfo {author} {\bibfnamefont {H.}~\bibnamefont {Strobel}}, \bibinfo
  {author} {\bibfnamefont {D.}~\bibnamefont {Linnemann}}, \bibinfo {author}
  {\bibfnamefont {A.}~\bibnamefont {Fr{\"o}lian}}, \bibinfo {author}
  {\bibfnamefont {T.}~\bibnamefont {Gasenzer}}, \bibinfo {author}
  {\bibfnamefont {M.}~\bibnamefont {Gärttner}},\ and\ \bibinfo {author}
  {\bibfnamefont {M.~K.}\ \bibnamefont {Oberthaler}},\ }\bibfield  {title}
  {\bibinfo {title} {{Spatially distributed multipartite entanglement enables
  EPR steering of atomic clouds}},\ }\href
  {https://doi.org/10.1126/science.aao2254} {\bibfield  {journal} {\bibinfo
  {journal} {Science}\ }\textbf {\bibinfo {volume} {360}},\ \bibinfo {pages}
  {413} (\bibinfo {year} {2018})}\BibitemShut {NoStop}%
\bibitem [{\citenamefont {Lange}\ \emph {et~al.}(2018)\citenamefont {Lange},
  \citenamefont {Peise}, \citenamefont {Lücke}, \citenamefont {Kruse},
  \citenamefont {Vitagliano}, \citenamefont {Apellaniz}, \citenamefont
  {Kleinmann}, \citenamefont {Tóth},\ and\ \citenamefont
  {Klempt}}]{Lange2018}%
  \BibitemOpen
  \bibfield  {author} {\bibinfo {author} {\bibfnamefont {K.}~\bibnamefont
  {Lange}}, \bibinfo {author} {\bibfnamefont {J.}~\bibnamefont {Peise}},
  \bibinfo {author} {\bibfnamefont {B.}~\bibnamefont {Lücke}}, \bibinfo
  {author} {\bibfnamefont {I.}~\bibnamefont {Kruse}}, \bibinfo {author}
  {\bibfnamefont {G.}~\bibnamefont {Vitagliano}}, \bibinfo {author}
  {\bibfnamefont {I.}~\bibnamefont {Apellaniz}}, \bibinfo {author}
  {\bibfnamefont {M.}~\bibnamefont {Kleinmann}}, \bibinfo {author}
  {\bibfnamefont {G.}~\bibnamefont {Tóth}},\ and\ \bibinfo {author}
  {\bibfnamefont {C.}~\bibnamefont {Klempt}},\ }\bibfield  {title} {\bibinfo
  {title} {{Entanglement between two spatially separated atomic modes}},\
  }\href {https://doi.org/10.1126/science.aao2035} {\bibfield  {journal}
  {\bibinfo  {journal} {Science}\ }\textbf {\bibinfo {volume} {360}},\ \bibinfo
  {pages} {416} (\bibinfo {year} {2018})}\BibitemShut {NoStop}%
\bibitem [{\citenamefont {Kunkel}\ \emph {et~al.}(2022)\citenamefont {Kunkel},
  \citenamefont {Pr\"ufer}, \citenamefont {Lannig}, \citenamefont {Strohmaier},
  \citenamefont {G\"arttner}, \citenamefont {Strobel},\ and\ \citenamefont
  {Oberthaler}}]{Kunkel2021}%
  \BibitemOpen
  \bibfield  {author} {\bibinfo {author} {\bibfnamefont {P.}~\bibnamefont
  {Kunkel}}, \bibinfo {author} {\bibfnamefont {M.}~\bibnamefont {Pr\"ufer}},
  \bibinfo {author} {\bibfnamefont {S.}~\bibnamefont {Lannig}}, \bibinfo
  {author} {\bibfnamefont {R.}~\bibnamefont {Strohmaier}}, \bibinfo {author}
  {\bibfnamefont {M.}~\bibnamefont {G\"arttner}}, \bibinfo {author}
  {\bibfnamefont {H.}~\bibnamefont {Strobel}},\ and\ \bibinfo {author}
  {\bibfnamefont {M.~K.}\ \bibnamefont {Oberthaler}},\ }\bibfield  {title}
  {\bibinfo {title} {{Detecting Entanglement Structure in Continuous Many-Body
  Quantum Systems}},\ }\href {https://doi.org/10.1103/PhysRevLett.128.020402}
  {\bibfield  {journal} {\bibinfo  {journal} {Phys. Rev. Lett.}\ }\textbf
  {\bibinfo {volume} {128}},\ \bibinfo {pages} {020402} (\bibinfo {year}
  {2022})}\BibitemShut {NoStop}%
\bibitem [{\citenamefont {Shchukin}\ and\ \citenamefont
  {Vogel}(2005)}]{Shchukin2005}%
  \BibitemOpen
  \bibfield  {author} {\bibinfo {author} {\bibfnamefont {E.}~\bibnamefont
  {Shchukin}}\ and\ \bibinfo {author} {\bibfnamefont {W.}~\bibnamefont
  {Vogel}},\ }\bibfield  {title} {\bibinfo {title} {{Inseparability Criteria
  for Continuous Bipartite Quantum States}},\ }\href
  {https://doi.org/10.1103/PhysRevLett.95.230502} {\bibfield  {journal}
  {\bibinfo  {journal} {Phys. Rev. Lett.}\ }\textbf {\bibinfo {volume} {95}},\
  \bibinfo {pages} {230502} (\bibinfo {year} {2005})}\BibitemShut {NoStop}%
\bibitem [{\citenamefont {Miranowicz}\ and\ \citenamefont
  {Piani}(2006)}]{Miranowicz2006}%
  \BibitemOpen
  \bibfield  {author} {\bibinfo {author} {\bibfnamefont {A.}~\bibnamefont
  {Miranowicz}}\ and\ \bibinfo {author} {\bibfnamefont {M.}~\bibnamefont
  {Piani}},\ }\bibfield  {title} {\bibinfo {title} {{Comment on
  ``Inseparability Criteria for Continuous Bipartite Quantum States''}},\
  }\href {https://doi.org/10.1103/PhysRevLett.97.058901} {\bibfield  {journal}
  {\bibinfo  {journal} {Phys. Rev. Lett.}\ }\textbf {\bibinfo {volume} {97}},\
  \bibinfo {pages} {058901} (\bibinfo {year} {2006})}\BibitemShut {NoStop}%
\bibitem [{\citenamefont {Miranowicz}\ \emph {et~al.}(2009)\citenamefont
  {Miranowicz}, \citenamefont {Piani}, \citenamefont {Horodecki},\ and\
  \citenamefont {Horodecki}}]{Miranowicz2009}%
  \BibitemOpen
  \bibfield  {author} {\bibinfo {author} {\bibfnamefont {A.}~\bibnamefont
  {Miranowicz}}, \bibinfo {author} {\bibfnamefont {M.}~\bibnamefont {Piani}},
  \bibinfo {author} {\bibfnamefont {P.}~\bibnamefont {Horodecki}},\ and\
  \bibinfo {author} {\bibfnamefont {R.}~\bibnamefont {Horodecki}},\ }\bibfield
  {title} {\bibinfo {title} {{Inseparability criteria based on matrices of
  moments}},\ }\href {https://doi.org/10.1103/PhysRevA.80.052303} {\bibfield
  {journal} {\bibinfo  {journal} {Phys. Rev. A}\ }\textbf {\bibinfo {volume}
  {80}},\ \bibinfo {pages} {052303} (\bibinfo {year} {2009})}\BibitemShut
  {NoStop}%
\bibitem [{\citenamefont {Opatrn\'y}\ and\ \citenamefont
  {Welsch}(1997)}]{Welsch1997}%
  \BibitemOpen
  \bibfield  {author} {\bibinfo {author} {\bibfnamefont {T.}~\bibnamefont
  {Opatrn\'y}}\ and\ \bibinfo {author} {\bibfnamefont {D.-G.}\ \bibnamefont
  {Welsch}},\ }\bibfield  {title} {\bibinfo {title} {{Density-matrix
  reconstruction by unbalanced homodyning}},\ }\href
  {https://doi.org/10.1103/PhysRevA.55.1462} {\bibfield  {journal} {\bibinfo
  {journal} {Phys. Rev. A}\ }\textbf {\bibinfo {volume} {55}},\ \bibinfo
  {pages} {1462} (\bibinfo {year} {1997})}\BibitemShut {NoStop}%
\bibitem [{\citenamefont {Mancini}\ \emph {et~al.}(1997)\citenamefont
  {Mancini}, \citenamefont {Tombesi},\ and\ \citenamefont
  {Man'ko}}]{Mancini1997}%
  \BibitemOpen
  \bibfield  {author} {\bibinfo {author} {\bibfnamefont {S.}~\bibnamefont
  {Mancini}}, \bibinfo {author} {\bibfnamefont {P.}~\bibnamefont {Tombesi}},\
  and\ \bibinfo {author} {\bibfnamefont {V.~I.}\ \bibnamefont {Man'ko}},\
  }\bibfield  {title} {\bibinfo {title} {{Density matrix from photon number
  tomography}},\ }\href {https://doi.org/10.1209/epl/i1997-00115-8} {\bibfield
  {journal} {\bibinfo  {journal} {EPL}\ }\textbf {\bibinfo {volume} {37}},\
  \bibinfo {pages} {79} (\bibinfo {year} {1997})}\BibitemShut {NoStop}%
\bibitem [{\citenamefont {Cramer}\ \emph {et~al.}(2010)\citenamefont {Cramer},
  \citenamefont {Plenio}, \citenamefont {Flammia}, \citenamefont {Somma},
  \citenamefont {Gross}, \citenamefont {Bartlett}, \citenamefont
  {Landon-Cardinal}, \citenamefont {Poulin},\ and\ \citenamefont
  {Liu}}]{Cramer2010}%
  \BibitemOpen
  \bibfield  {author} {\bibinfo {author} {\bibfnamefont {M.}~\bibnamefont
  {Cramer}}, \bibinfo {author} {\bibfnamefont {M.~B.}\ \bibnamefont {Plenio}},
  \bibinfo {author} {\bibfnamefont {S.~T.}\ \bibnamefont {Flammia}}, \bibinfo
  {author} {\bibfnamefont {R.}~\bibnamefont {Somma}}, \bibinfo {author}
  {\bibfnamefont {D.}~\bibnamefont {Gross}}, \bibinfo {author} {\bibfnamefont
  {S.~D.}\ \bibnamefont {Bartlett}}, \bibinfo {author} {\bibfnamefont
  {O.}~\bibnamefont {Landon-Cardinal}}, \bibinfo {author} {\bibfnamefont
  {D.}~\bibnamefont {Poulin}},\ and\ \bibinfo {author} {\bibfnamefont {Y.-K.}\
  \bibnamefont {Liu}},\ }\bibfield  {title} {\bibinfo {title} {{Efficient
  quantum state tomography}},\ }\href {https://doi.org/10.1038/ncomms1147}
  {\bibfield  {journal} {\bibinfo  {journal} {Nat. Comm.}\ }\textbf {\bibinfo
  {volume} {1}},\ \bibinfo {pages} {149} (\bibinfo {year} {2010})}\BibitemShut
  {NoStop}%
\bibitem [{\citenamefont {Hertz}\ \emph {et~al.}(2019)\citenamefont {Hertz},
  \citenamefont {Oreshkov},\ and\ \citenamefont {Cerf}}]{Hertz2019b}%
  \BibitemOpen
  \bibfield  {author} {\bibinfo {author} {\bibfnamefont {A.}~\bibnamefont
  {Hertz}}, \bibinfo {author} {\bibfnamefont {O.}~\bibnamefont {Oreshkov}},\
  and\ \bibinfo {author} {\bibfnamefont {N.~J.}\ \bibnamefont {Cerf}},\
  }\bibfield  {title} {\bibinfo {title} {{Multicopy uncertainty observable
  inducing a symplectic-invariant uncertainty relation in position and momentum
  phase space}},\ }\href {https://doi.org/10.1103/PhysRevA.100.052112}
  {\bibfield  {journal} {\bibinfo  {journal} {Phys. Rev. A}\ }\textbf {\bibinfo
  {volume} {100}},\ \bibinfo {pages} {052112} (\bibinfo {year}
  {2019})}\BibitemShut {NoStop}%
\bibitem [{\citenamefont {Arnhem}\ \emph {et~al.}(2022)\citenamefont {Arnhem},
  \citenamefont {Griffet},\ and\ \citenamefont {Cerf}}]{Griffet2022a}%
  \BibitemOpen
  \bibfield  {author} {\bibinfo {author} {\bibfnamefont {M.}~\bibnamefont
  {Arnhem}}, \bibinfo {author} {\bibfnamefont {C.}~\bibnamefont {Griffet}},\
  and\ \bibinfo {author} {\bibfnamefont {N.~J.}\ \bibnamefont {Cerf}},\
  }\bibfield  {title} {\bibinfo {title} {{Multicopy observables for the
  detection of optically nonclassical states}},\ }\href
  {https://doi.org/10.1103/PhysRevA.106.043705} {\bibfield  {journal} {\bibinfo
   {journal} {Phys. Rev. A}\ }\textbf {\bibinfo {volume} {106}},\ \bibinfo
  {pages} {043705} (\bibinfo {year} {2022})}\BibitemShut {NoStop}%
\bibitem [{\citenamefont {Griffet}\ \emph {et~al.}(2022)\citenamefont
  {Griffet}, \citenamefont {Arnhem}, \citenamefont {Bièvre},\ and\
  \citenamefont {Cerf}}]{Griffet2022b}%
  \BibitemOpen
  \bibfield  {author} {\bibinfo {author} {\bibfnamefont {C.}~\bibnamefont
  {Griffet}}, \bibinfo {author} {\bibfnamefont {M.}~\bibnamefont {Arnhem}},
  \bibinfo {author} {\bibfnamefont {S.~D.}\ \bibnamefont {Bièvre}},\ and\
  \bibinfo {author} {\bibfnamefont {N.~J.}\ \bibnamefont {Cerf}},\ }\bibfield
  {title} {\bibinfo {title} {{Interferometric measurement of the quadrature
  coherence scale using two replicas of a quantum optical state}},\ }\href@noop
  {} {\bibfield  {journal} {\bibinfo  {journal}
  {\href{https://arxiv.org/abs/2211.12992}{arXiv:2211.12992}}\ } (\bibinfo
  {year} {2022})}\BibitemShut {NoStop}%
\bibitem [{\citenamefont {Brun}(2004)}]{Brun2004}%
  \BibitemOpen
  \bibfield  {author} {\bibinfo {author} {\bibfnamefont {T.~A.}\ \bibnamefont
  {Brun}},\ }\bibfield  {title} {\bibinfo {title} {{Measuring Polynomial
  Functions of States}},\ }\href@noop {} {\bibfield  {journal} {\bibinfo
  {journal} {Quantum Inf. Comput.}\ }\textbf {\bibinfo {volume} {4}},\ \bibinfo
  {pages} {401–408} (\bibinfo {year} {2004})}\BibitemShut {NoStop}%
\bibitem [{\citenamefont {Einstein}\ \emph {et~al.}(1935)\citenamefont
  {Einstein}, \citenamefont {Podolsky},\ and\ \citenamefont
  {Rosen}}]{Einstein1935}%
  \BibitemOpen
  \bibfield  {author} {\bibinfo {author} {\bibfnamefont {A.}~\bibnamefont
  {Einstein}}, \bibinfo {author} {\bibfnamefont {B.}~\bibnamefont {Podolsky}},\
  and\ \bibinfo {author} {\bibfnamefont {N.}~\bibnamefont {Rosen}},\ }\bibfield
   {title} {\bibinfo {title} {{Can Quantum-Mechanical Description of Physical
  Reality Be Considered Complete?}},\ }\href
  {https://doi.org/10.1103/PhysRev.47.777} {\bibfield  {journal} {\bibinfo
  {journal} {Phys. Rev.}\ }\textbf {\bibinfo {volume} {47}},\ \bibinfo {pages}
  {777} (\bibinfo {year} {1935})}\BibitemShut {NoStop}%
\bibitem [{\citenamefont {Sanders}(1989)}]{Sanders1989}%
  \BibitemOpen
  \bibfield  {author} {\bibinfo {author} {\bibfnamefont {B.~C.}\ \bibnamefont
  {Sanders}},\ }\bibfield  {title} {\bibinfo {title} {{Quantum dynamics of the
  nonlinear rotator and the effects of continual spin measurement}},\ }\href
  {https://doi.org/10.1103/PhysRevA.40.2417} {\bibfield  {journal} {\bibinfo
  {journal} {Phys. Rev. A}\ }\textbf {\bibinfo {volume} {40}},\ \bibinfo
  {pages} {2417} (\bibinfo {year} {1989})}\BibitemShut {NoStop}%
\bibitem [{\citenamefont {Goldberg}\ \emph {et~al.}(2023)\citenamefont
  {Goldberg}, \citenamefont {Thekkadath},\ and\ \citenamefont
  {Heshami}}]{Goldberg2023}%
  \BibitemOpen
  \bibfield  {author} {\bibinfo {author} {\bibfnamefont {A.~Z.}\ \bibnamefont
  {Goldberg}}, \bibinfo {author} {\bibfnamefont {G.~S.}\ \bibnamefont
  {Thekkadath}},\ and\ \bibinfo {author} {\bibfnamefont {K.}~\bibnamefont
  {Heshami}},\ }\bibfield  {title} {\bibinfo {title} {Measuring the quadrature
  coherence scale on a cloud quantum computer},\ }\href
  {https://doi.org/10.1103/PhysRevA.107.042610} {\bibfield  {journal} {\bibinfo
   {journal} {Physical Review A}\ }\textbf {\bibinfo {volume} {107}},\ \bibinfo
  {pages} {042610} (\bibinfo {year} {2023})}\BibitemShut {NoStop}%
\end{thebibliography}%

\end{document}